\documentclass[prb,preprint,amsmath,amssymb, superscriptaddress]{revtex4-2}

\usepackage{textcomp}
\usepackage{amsmath,graphicx}
\usepackage{blindtext}
\usepackage{dcolumn}
\usepackage{tabularx}
\usepackage{physics}
\usepackage{hyperref}
\usepackage[capitalise]{cleveref}
\usepackage{multirow}
\usepackage{xcolor}

\newcommand{\Rn}{\ensuremath{R_{n}}}
\newcommand{\didv}{\ensuremath{\mathrm{d}I/\mathrm{d}V}}

\begin{document}

\title{Multiple Andreev Reflection Effects in Asymmetric STM Josephson Junctions}

\author{Wan-Ting Liao}
\affiliation{Laboratory for Physical Sciences, College Park, Maryland 20740}
\affiliation{Quantum Materials Center, Department of Physics, University of Maryland, College Park, Maryland 20742}
\author{S. K. Dutta}
\affiliation{Quantum Materials Center, Department of Physics, University of Maryland, College Park, Maryland 20742}
\author{R. E. Butera}
\affiliation{Laboratory for Physical Sciences, College Park, Maryland 20740}
\author{C. J. Lobb}
\affiliation{Quantum Materials Center, Department of Physics, University of Maryland, College Park, Maryland 20742}
\affiliation{Joint Quantum Institute, University of Maryland, College Park, Maryland 20742}
\author{F. C. Wellstood}
\affiliation{Quantum Materials Center, Department of Physics, University of Maryland, College Park, Maryland 20742}
\affiliation{Joint Quantum Institute, University of Maryland, College Park, Maryland 20742}
\author{M. Dreyer}
\affiliation{Laboratory for Physical Sciences, College Park, Maryland 20740}
\date{\today}

\begin{abstract}
We have examined the electrical behavior of Josephson junctions formed by a scanning tunneling microscope (STM) with a Nb sample and a Nb tip, with normal-state resistances $R_n$ varying between 1 k$\Omega$ and 10 M$\Omega$. Current-voltage characteristics were obtained as a function of $R_n$ by varying the distance between the tip and sample at temperatures of 50 mK and 1.5 K. $R_n$ decreases as the tip-sample separation is reduced, and the junction evolves from a phase-diffusion regime to an underdamped small junction regime, and then to a point contact regime. The subgap structure exhibits pronounced multiple Andreev reflection (MAR) features whose amplitudes and onset energies depend sensitively on junction transparency and gap asymmetry. To interpret these spectra, we generalize the Averin–Bardas MAR theory to superconductors with unequal gap magnitudes, providing a quantitative model appropriate for asymmetric STM junctions. The resulting fits yield the superconducting gaps of the electrodes, barrier transparency, and number of conduction channels as a function of $R_n$. Combining this analysis with Josephson junction dynamics, we further account for the observed switching and retrapping currents and the finite resistance of the supercurrent branch. Our results demonstrate that incorporating intrinsic electrode asymmetry is essential for reliably extracting transport parameters in STM-based superconducting weak links.
\end{abstract}

\maketitle
\section{\label{sec:level1}Introduction}

When electrical current flows freely across a good metallic contact, the voltage drop reflects the normal resistivity of the materials. In contrast, when two conductors are connected through a "weak link," the coupling between them is limited by the transmission through a small number of electronic channels. For example, when the weak link couples two superconductors, a Josephson junction is formed \cite{JOSEPHSON1962251}. In the pioneering work of Scheer \textit{et al.} \cite{scheer1998signature}, atomic-scale superconducting break-junctions were fabricated using an STM (scanning tunneling microscope) technique \cite{agrait1992transition, rodrigo2004superconducting, guillamon2008scanning, rodrigo2006josephson} or by controllably bending a thin film, forming symmetric Josephson junctions. The transport properties of such junctions are governed by the intrinsic properties of the electrodes and the microscopic details of the coupling region, giving rise to phenomena such as Josephson supercurrents and multiple Andreev reflection (MAR) \cite{averin1995ac, cuevas1996hamiltonian, scheer1998signature, MULLER1992485, scheer1997conduction, PhysRevB.74.132501, rodrigo2004superconducting, agrait2003quantum, hiraoka2014transport}. Through MAR spectroscopy, they observed individual channels linked to the atomic chemical valence of the material, establishing MAR as a powerful probe of superconducting transport. 

A Josephson STM \cite{PhysRevLett.101.037002,Hamidian2016,PhysRevB.93.161115,ast2016sensing,PhysRevB.93.020504, jack2017quantum, senkpiel2020single, PhysRevLett.124.156803,PhysRevResearch.6.043233,PhysRevLett.123.017001} provides a complementary platform for investigating atomic scale superconducting junctions. A key advantage of this platform is the ability to tune the weak-link properties in situ by varying the tip–sample separation, thereby controlling the normal-state resistance $R_n$. Recently, MAR theory combined with $P(E)$ theory has enabled a quantitative interpretation of Josephson STM measurements, relating subgap transport and Josephson currents to microscopic channel transmission \cite{PhysRevResearch.6.043233,senkpiel2020single}. In Josephson STM junctions, asymmetry in superconducting energy gaps and electrode geometries is common \cite{PhysRevB.74.132501,PhysRevApplied.4.034011}. Such asymmetry can arise from the use of different STM tip and sample materials, or from preparation-induced gap variations, even for nominally identical electrodes. Similar gap mismatches are also common in other superconducting devices, such as those used for quantum computation \cite{steffen2023characterization}.   

To understand how MAR manifests in junctions with asymmetric superconducting gaps and electrode geometries, we examined how spectroscopic features in their current-voltage $I-V$ characteristics depend on the properties of the weak link. In this work, we used a cryogenic STM \cite{roychowdhury201430} with a Nb tip and Nb sample to form superconducting weak links. We varied the junction normal state resistances $R_n$ between 1 k$\Omega$ to 10 M$\Omega$, enabling both voltage- and current-biased measurements of MAR subgap features and other key junction parameters such as the switching current $I_s$, retrapping current $I_r$, and supercurrent branch resistance $R_z$. Depending on $R_n$, the junction operates in the phase-diffusion regime \cite{PhysRevLett.101.037002,proslier2006probing,PhysRevB.78.140507}, the underdamped small-junction limit \cite{jack2017quantum,trahms2023diode,lu2023phase}, or the point-contact regime \cite{rodrigo2004superconducting,agrait1992transition}.

We analyzed the resulting $I-V$ characteristics using a generalized Averin–Bardas framework \cite{averin1995ac} for the case of superconductors with different gap magnitudes, yielding an asymmetric MAR model applicable to STM junctions. By fitting to the measured $I-V$ and $\didv$ characteristics, we obtained the transparency, superconducting gaps, and number of conduction channels as a function of junction resistance. We find that superconducting gap asymmetry is essential for accurately capturing the evolution of the subgap MAR structure with increasing transparency. Additional measurements allow us to explore the crossover between different transport regimes and the role of phase diffusion in small Josephson junctions.


\section{\label{sec:level2}Theory}

MAR has been theoretically addressed mainly by two different approaches. The first, from Averin and Bardas (AB), characterizes the current in a weak link constriction by a scattering matrix \cite{averin1995ac}. The second approach, from Cuevas \textit{et al.}, uses a Hamiltonian of the total current through the contact, solved with nonequilibrium Green's function techniques \cite{cuevas1996hamiltonian}. Both assumed a symmetric junction, composed of two electrodes with the same superconducting gap.  Later, Wu \textit{et al.} \cite{wu2004ac} generalized the Hamiltonian approach for asymmetric junctions. In this section, we adapt the theoretical AB approach, which provides a vivid physical picture of the processes involved, to the case of superconductors with different gaps. 

\begin {figure}[t]
\centering
\includegraphics[width=0.9\linewidth]{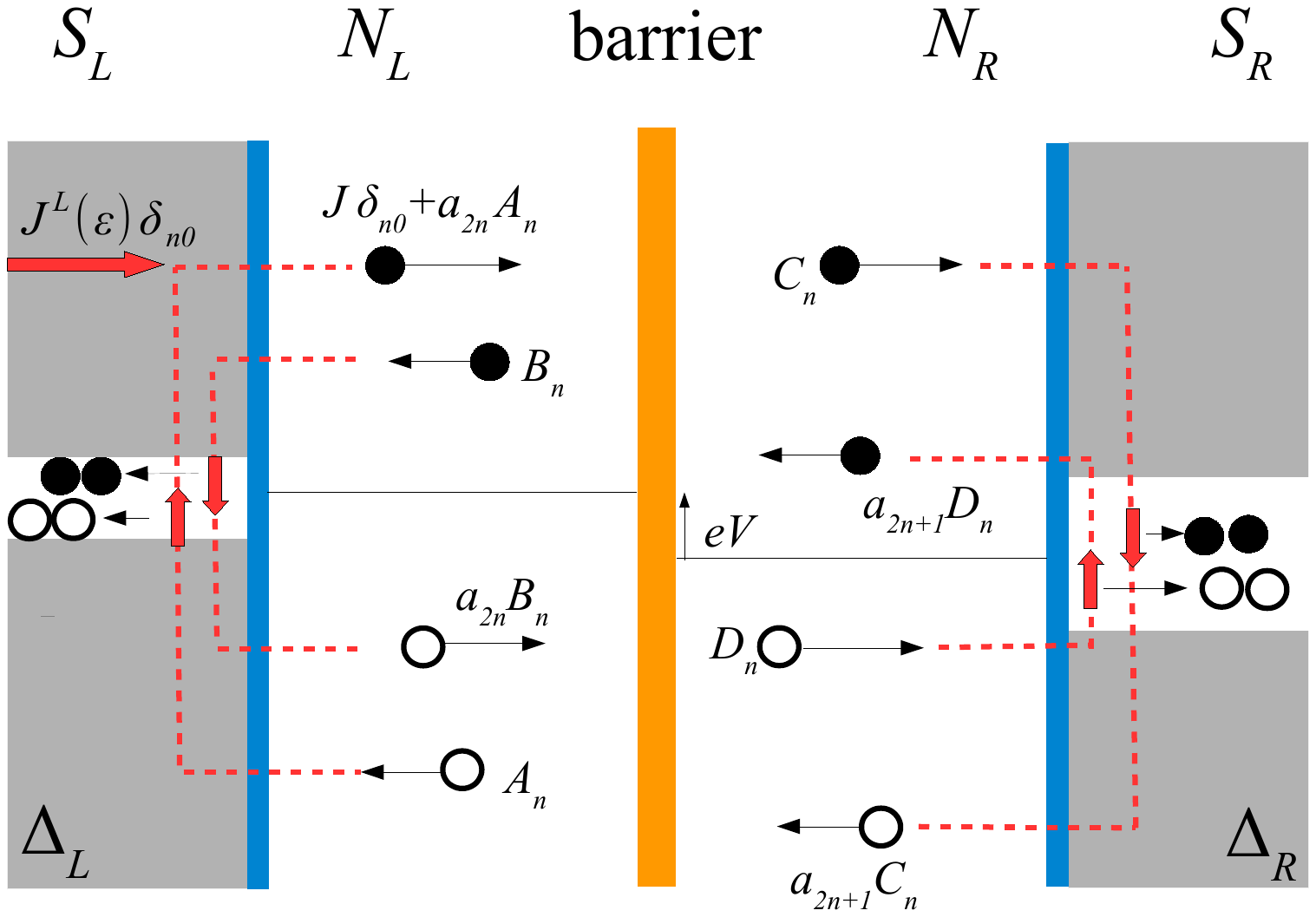}
\caption{Illustration showing MAR wavefunction components of the generalized AB model for the case of electron-like source quasiparticles incident from the left. The left lead is a superconductor with gap $\Delta_{L}$ and the right lead is a superconductor with gap $\Delta_{R}$. A voltage bias $V$ is applied to the right lead with
respect to the left lead. The barrier in the center acts as a scattering region for quasiparticles and is characterized by transparency $D$ (distinct from the incident hole term $D_n$). The superconductor-to-normal metal interfaces (blue) have unity transparency.}
\label{averin}
\end{figure}

\Cref{averin} illustrates a generalized AB model for asymmetric MAR, which allows for different superconducting gaps in the two leads. The left lead is superconductor $S_{L}$ with gap $\Delta_{L}$ and the right lead is superconductor $S_{R}$ with gap $\Delta_{R}$. A voltage bias $V$ is applied to the right lead with respect to the left. Between the two superconductors are normal metal regions, $N_{L}$ and $N_{R}$, separated by a scattering potential that sets the barrier transparency $D$, the probability that an incident particle is transmitted.  All electron and hole scattering effects in the normal metal are incorporated into $D$ and we assume the interface between each superconductor and the normal region has unity transparency. In the absence of scattering, the link is completely transparent and $D=1$. Otherwise, $D$ falls between 0 and 1. An electron in $N_{L}$ that is incident on $S_{L}$, for example, may be Andreev reflected as a hole (as shown by the red dashed line), generating a Cooper pair in $S_{L}$. An analogous process occurs for an incident hole. Similarly, an electron (hole) in $N_{R}$ incident on $S_{R}$ may be Andreev reflected as an hole (electron).
 
Multiple Andreev reflections can occur because an electron traveling to the right in region $N_{L}$ may pass through the barrier (with a certain probability amplitude) and then be Andreev reflected at the $N_{R}$-$S_{R}$ interface, emerging as a hole in $N_{R}$. The reflected hole may then travel back through the barrier and arrive at the $N_{L}$-$S_{L}$ interface where it can again be Andreev reflected with some probability amplitude. In the AB analysis, this process continues, allowing the iterative construction of the electron and hole wavefunctions in the normal regions. The wavefunctions in $N_{L}$ and $N_{R}$ are interdependent because the net current must be the same in both regions. The detailed nature of this dependence is determined by the barrier, and can be represented by a scattering matrix. With knowledge of the scattering matrix, recursion relations are derived to obtain the wavefunction coefficients \cite{averin1995ac}. A detailed schematic of the Andreev process can be found in Fig.~S1 in the Supplemental Material \cite{ref1}.

When $S_{L}$ and $S_{R}$ have different gaps, then the Andreev reflection amplitudes will not depend solely on the energy relative to the Fermi energy, but also on which interface is reflecting the particle. For the $N_{L}$-$S_{L}$ interface, the Andreev reflection amplitude of a quasiparticle with energy $E$ is \cite{averin1995ac}
\begin{equation}
   a^{L}(E) = \frac{1}{\Delta_{L}} \times \left\{\begin{array}{lr}
       E-\textrm{sgn}(E)(E^2-\Delta_{L}^2)^{1/2} , & |E|>\Delta_{L}\\
        E-i(\Delta_{L}^2-E^2)^{1/2} , &|E|<\Delta_{L}\\    
        \end{array}\right.
        \label{eq:ch71:aepsilonl}
\end{equation}  
and for the the right interface, it is
\begin{equation}
   a^{R}(E) = \frac{1}{\Delta_{R}} \times \left\{\begin{array}{lr}
        E-\textrm{sgn}(E)(E^2-\Delta_{R}^2)^{1/2} , & |E|>\Delta_{R}\\
        E-i(\Delta_{R}^2-E^2)^{1/2}, &  |E|<\Delta_{R}.\\    
        \end{array}\right. 
        \label{eq:ch71:aepsilonr}
\end{equation}
 
Because the Andreev reflection amplitudes are different at the left and right $N-S$ interfaces, it is necessary to explicitly consider hole-like quasiparticles incident from the left and right electrodes, as well as electron-like quasiparticles incident from both electrodes, for a total of four separate processes. To calculate the total current, we find the corresponding wavefunctions and sum the current from each process. 

In region $N_{L}$, the wavefunctions that result from an electron-like quasiparticle incident from $S_{L}$ are
\begin{widetext}
\begin{align}
   \psi_{e}^{eL}&=\displaystyle\sum_{n=-\infty}^{\infty} [(a_{\{2n,+ \}}^{L} A_n^{\rightarrow}+J^{L} \delta_{n0})e^{+ ikx}+B_n^{\rightarrow}e^{- ikx}]e^{-i(\varepsilon+ 2neV)t/\hbar}\nonumber\\
    \psi_{h}^{eL}&=\displaystyle\sum_{n=-\infty}^{\infty} [A_n^{\rightarrow}e^{+ ikx}+a_{\{2n,+ \}}^{L} B_n^{\rightarrow} e^{- ikx}]e^{-i(\varepsilon+ 2neV)t/\hbar},
    \label{eq:el} 
\end{align}
\end{widetext}
where $\psi_{e}^{eL}$ and $\psi_{h}^{eL}$ represent electrons and holes, respectively. $J^{L}(\varepsilon)=\sqrt{1-|a^{L}(\varepsilon)|^2}$ is the quasiparticle source current term from $S_{L}$. $k$ and $\varepsilon$ are the Fermi wavevector and energy of the incident quasiparticle, and $t$ is the time. $a_{\{2n,+ \}}^{L}=a^{L}(\varepsilon+2neV)$ is the Andreev reflection amplitude. The wavefunction amplitudes shown in \cref{averin} only apply to the case of an electron-like quasiparticle source incident from the left, with further details given in the Supplemental Material \cite{ref1}. 

Here, an electron acquires a total energy of $2eV$ during a cycle in which it travels from $S
_L$ to $S_R$ through the barrier, undergoes Andreev reflection as a hole at the $N_R-S_R$ interface, propagates back through the barrier, and is finally Andreev reflected back into an electron at the $N_L - S_L$ interface. $A_n^{\rightarrow}$ and $B_n^{\rightarrow}$ are the amplitudes of leftward moving holes and electrons in region $N_{L}$ that result from the sum of reflecting off the barrier and transmitting through the barrier from the right interface, at order $n$. We use the arrow label over $A_n$ and $B_n$ to signify that these amplitudes came from source quasiparticles incident from the left. 

Analogously to \cref{eq:el}, in region $N_{L}$ the wavefunctions produced by a hole-like quasiparticle incident from $S_{L}$ are
\begin{widetext}
\begin{align}
   \psi_{h}^{hL}&=\displaystyle\sum_{n=-\infty}^{\infty} [(a_{\{2n,- \}}^{L} A_n^{\rightarrow}+J^{L} \delta_{n0})e^{- ikx}+B_n^{\rightarrow}e^{+ ikx}]e^{ -i(\varepsilon- 2neV)t/\hbar}\nonumber\\
    \psi_{e}^{hL}&=\displaystyle\sum_{n=-\infty}^{\infty} [A_n^{\rightarrow}e^{-ikx}+a_{\{2n,- \}}^{L} B_n^{\rightarrow} e^{+ ikx}]e^{-i(\varepsilon - 2neV)t/\hbar}.
    \label{eq:hl}
\end{align} 
\end{widetext}
$\psi_{h}^{hL}$ and $\psi_{e}^{hL}$ represent the wavefunctions of the holes and electrons, respectively, for a hole-like quasiparticle incident from the left. $a_{\{2n,- \}}^{L}=a^{L}(\varepsilon-2neV)$ denotes the Andreev reflection amplitude. The minus sign in the energy argument arises because a hole incident from the left loses kinetic energy $eV$ when moving from $S_{L}$ to $S_{R}$. In this case, $A_n^{\rightarrow}$ and $B_n^{\rightarrow}$ are the amplitudes of leftward moving electrons and holes in region $N_{L}$.    

The total probability current density is then
\begin{widetext}
\begin{align}
j_{L}
&=\frac{\hbar}{2mi}[(\psi_{e}^{eL*} \grad\psi_{e}^{eL}-\psi_{e}^{eL}\grad \psi_{e}^{eL*})+(\psi_{h}^{eL*} \grad\psi_{h}^{eL}-\psi_{h}^{eL}\grad \psi_{h}^{eL*})] \nonumber\\
&+\frac{\hbar}{2mi}[(\psi_{e}^{hL*} \grad\psi_{e}^{hL}-\psi_{e}^{hL}\grad \psi_{e}^{hL*})+(\psi_{h}^{hL*} \grad\psi_{h}^{hL}-\psi_{h}^{hL}\grad \psi_{h}^{hL*})],
    \label{eq:sis:wvd2}
\end{align}
\end{widetext}
where $m$ is the mass of the quasiparticle.

The total electrical current $I_L$ due to an electron and hole incident from the left is directly proportional to this current density. Summing over the source energy gives
\begin{align}
I_{L}(t)&=\frac{e}{2\pi \hbar}\frac{m}{\hbar k} \int_{-\infty}^{\infty}d\varepsilon  j_{L}(\varepsilon,t) f(\varepsilon), \nonumber\\
\label{eq:sis:ctd2e}
\end{align}
where $f(\varepsilon$)=$(e^{\varepsilon/k_BT}+1)^{-1}$ is the electron occupancy for states at energy $\varepsilon$, which we assume is a Fermi distribution. 

This completes half of the story. To calculate the total current $I(t)$, we also need to consider the current $I_{R}(t)$ due to an electron-like or hole-like quasiparticle incident from the right. The main physics remains the same, except we need to take into account that (1) an electron-like quasiparticle loses kinetic energy $eV$ when traveling from the right to the left interface and (2) a hole-like quasiparticle gains energy $eV$ when traveling from the right to the left interface. Therefore, the wavefunctions $\psi_{e}^{eR}$ and $\psi_{h}^{eR}$ for an electron incident from the right can be found from $\psi_{h}^{hL}$ and $\psi_{e}^{hL}$ by substituting superscript $R$ for $L$, and $\leftarrow$ for $\rightarrow$. Similarly the wavefunctions $\psi_{h}^{hR}$ and $\psi_{e}^{hR}$ can be found from $\psi_{e}^{eL}$ and $\psi_{h}^{eL}$ with the same substitutions. Full expressions for these wavefunctions are given in the Supplemental Material \cite{ref1}.

Combining the left and right currents, the total junction current is 
\begin{align}
I_{L}(t)+I_{R}(t)&=\sum_{k}^{} I_ke^{i2keVt/\hbar}, \hspace{8em} \nonumber\\\label{eq:sis:cthe41}
\end{align}

where
\begin{widetext}
\begin{align}
I_k&=\frac{e}{2\pi \hbar}\bigg[2eVD\delta_{k0}-\int_{-\infty}^{\infty}d\varepsilon \tanh({\frac{\varepsilon}{2k_{B}T}})\bigg(J^{L}(\varepsilon)[(a^{  \rightarrow}_{2k})^* (A^{\rightarrow}_{k})^* +a^{\rightarrow}_{-2k} A^{\rightarrow}_{-k} ] \nonumber\\
& + \sum_{n}[1+a^{\rightarrow}_{2n} (a^{\rightarrow}_{2(n+k)})^* ][A^{\rightarrow}_n (A^{ \rightarrow}_{n+k})^*  -B_n^{\rightarrow}(B_{n+k}^{\rightarrow})^* \big]\bigg)+\int_{-\infty}^{\infty}d\varepsilon \tanh({\frac{\varepsilon}{2k_{B}T}})\nonumber\\
&\bigg(J^{R}(\varepsilon)[a_{2k}^{\leftarrow} A_{k}^{\leftarrow} +(a_{-2k}^{\leftarrow})^* (A_{-k}^{\leftarrow})^* ] + \sum_{n}[1+(a_{2n}^{\leftarrow})^* a_{2(n+k)}^{\leftarrow} ][(A_{n}^{\leftarrow})^* A_{n+k}^{\leftarrow} -(B_{n}^{\leftarrow})^* B_{n+k}^{\leftarrow} ] \bigg)\bigg].
\label{eq:sis:cthe4}
\end{align}
\end{widetext}

Here we have simplified the sums by defining $a_{n}^{\rightarrow}=a^{R}(\varepsilon+neV)$ for $n$ odd and $a_{n}^{\rightarrow}=a^{L}(\varepsilon+neV)$ for $n$ even. Thus, an odd valued subscript corresponds to a component that had an Andreev reflection from $S_{R}$ and an even subscript to Andreev reflection from $S_{L}$. Similarly $a_{n}^{\leftarrow} =a^{L}(\varepsilon-neV)$ for $n$ odd and $a_{n}^{\leftarrow} =a^{R}(\varepsilon-neV)$ for $n$ even; the odd indices are for $S_{L}$ and the even indices are for $S_{R}$.

To calculate the $I-V$ characteristic, we need to know the amplitudes $A_n^{\rightarrow}$, $B_n^{\rightarrow}$, $A_n^{\leftarrow}$, and $B_n^{\leftarrow}$. Following AB's approach \cite{averin1995ac}, we let $C_n^{\rightarrow}$ and $D_n^{\rightarrow}$ denote the wavefunction amplitudes in $N_R$ (shown in \cref{averin}). Relating the incoming electrons to the outgoing electrons in regions $N_{L}$ and $N_{R}$ using a scattering matrix $S_e$ where
\begin{align}
S_{e}=
 \begin{pmatrix}
  r & t  \\
  t & -r^*t/t^* 
 \end{pmatrix},
 \label{eq:sis:scatmatrix}
\end{align}
we obtain a matrix equation for electrons. Here, $r$ and $t$ are the normal-state reflection and transmission amplitudes of the junction at energy $\varepsilon$. The transparency is $D=|t|^2$. Similarly, there will be a matrix equation with a scattering matrix $S_{h}=S_{e}^*$ connecting the holes in the two regions. 

After solving two matrix equations for electrons and holes (detailed derivation can be found in Ref.~\cite{wtthesis}), the wavefunction amplitudes $A_n^{\rightarrow}$ and $B_n^{\rightarrow}$ are
\begin{widetext}
\begin{align}
 A_{n+1}^{\rightarrow}-a_{2n+1}^{R}a_{2n}^{L}A_n^{\rightarrow}=\sqrt{R}(a_{2n+2}^{L} B_{n+1}^{\rightarrow}-a_{2n+1}^{R} B_n^{\rightarrow})+J^La_1^{R}\delta_{n0} 
 \label{eq:sis:nrec1}
\end{align}
\begin{align}
D\frac{a_{2n+1}^{R} a_{2n+2}^{L}}{1-(a_{2n+1}^{L})^2}B_{n+1}^{\rightarrow}&-\bigg[D\bigg(\frac{(a_{2n+1}^{R})^2}{1-(a_{2n+1}^{R})^2}+\frac{(a_{2n}^{L})^2}{1-(a_{2n-1}^{R})^2}\bigg)+1-(a_{2n}^{L})^2\bigg]B_n^{\rightarrow}\nonumber\\
&+D\frac{a_{2n}^{L}a_{2n-1}^{R}}{1-a_{2n-1}^{R}}B_{n-1}^{\rightarrow}=-\sqrt{R}J^L\delta_{n0},
\label{eq:sis:nrec2}
\end{align}
\end{widetext}
where $R=1-D$ is the reflection coefficient of the barrier in the normal region. The same recursion relations describe hole-like source quasiparticles incident from the left superconductor.  

Similarly, to obtain the wavefunction amplitudes $A_n^{\leftarrow}$ and $B_n^{\leftarrow}$ for electron-like and hole-like source quasiparticles incident from the right superconductor, we swap $L$ for $R$ and $\leftarrow$ for $\rightarrow$ in Eqs.~\eqref{eq:sis:nrec1} and \eqref{eq:sis:nrec2}, by symmetry. We then plug these into Eq.~\eqref{eq:sis:cthe4} and obtain the Fourier components $I_k$ using numerical integration.  $I_0$ represents the dc current, while the junction critical current is given by \cite{averin1995ac}
\begin{align}
I_c=\lim_{V \to 0} 2|\operatorname{Im}(I_1)|.
\label{ic}
\end{align}
These currents will depend on the transparency of the junction and the superconducting gap of its two electrodes. 

\begin {figure}[t]
\centering
\includegraphics[width=1.0\linewidth]{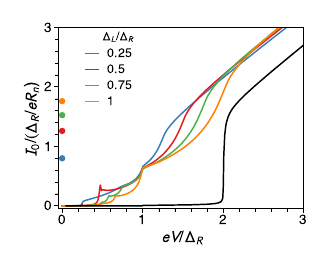}
\caption{$I-V$ characteristics for superconducting gap ratios ${\Delta}_{L}$/${\Delta}_{R}=0.25$ (blue), $0.5$ (red), $0.75$ (green) and $1$ (orange) at transparency $D=0.4$ for $T=0$ K. Circles at $V=0$ show corresponding critical currents from \cref{ic}. For comparison, also shown is the $I-V$ curve for a symmetric junction with $D=0.01$ (black), near the conventional tunnel limit.}
\label{ratio}
\end{figure}

\Cref{ratio} shows dc $I-V$ characteristics calculated from \cref{eq:sis:cthe4} for $D=0.4$ with different superconducting gap ratios ${\Delta}_{L}/{\Delta}_{R}$. In the plot, the current is normalized by a factor that includes the normal-state resistance $R_n = 1/(G_0 D)$, where $G_0 = e^2/(\pi \hbar) = 1 / (12.9\ \textrm{k}\Omega)$ is the conductance quantum. Our calculations appear to agree with previous results for ${\Delta}_{L}/{\Delta}_{R}=1$ \cite{averin1995ac} and 0.5 \cite{wu2004ac}. The $I-V$ curves show a large subgap current compared to the low transparency tunneling limit, with steps at energies $eV$ near ${\Delta}_{L}/n$, ${\Delta}_{R}/n$, and $({\Delta}_{L}+{\Delta}_{R})/n$ for integral values of $n$.

\begin {figure}[t]
\centering
\includegraphics[width=1.0\linewidth]{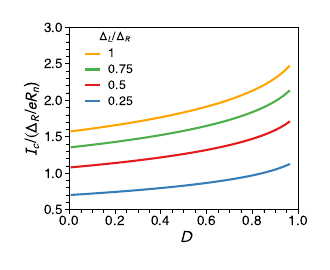}
\caption{Normalized critical current as a function of $D$ for ${\Delta}_{L}$/${\Delta}_{R}=0.25$ (blue), $0.5$ (red), $0.75$ (green) and $1$ (orange) at $T=0$ K. }
\label{d}
\end{figure}

\begin {figure}[t]
\centering
\includegraphics[width=1\linewidth]{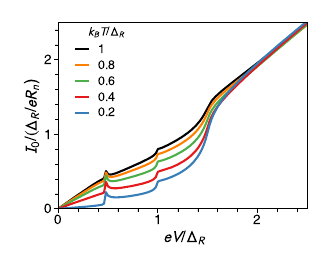}
\caption{Effects of temperature on calculated $I-V$ characteristics for transparency $D=0.25$ and ${\Delta}_{L}$/${\Delta}_{R}=0.5$. The curves correspond to $k_{B}T/{\Delta}_{R}$ values of 1 (black), 0.8 (orange), 0.6 (green), 0.4 (red), and 0.2 (blue).}
\label{temp}
\end{figure}

\Cref{d} shows the normalized critical current calculated from \cref{ic} as a function of the transparency $D$ for four values of the gap ratio ${\Delta}_{L}/{\Delta}_{R}$. As expected, the critical current becomes larger for larger ${\Delta}_{L}+{\Delta}_{R}$ and increases with $D$. For symmetric superconducting gaps, the critical current approaches the Ambegaokar-Baratoff formula $I_c R_n=\pi\Delta/2e$ as $D\rightarrow0$ \cite{PhysRevLett.10.486}. \Cref{temp} shows examples of dc $I-V$ characteristics at different temperatures, for $D$ = 0.25 and $\Delta_L/\Delta_R = 0.5$. At high temperature, the subgap current increases and shows thermally-rounded MAR steps.

\section{\label{sec:level3}Experimental methods}  

\begin {figure}[t]
\centering
\includegraphics[width=1\linewidth]{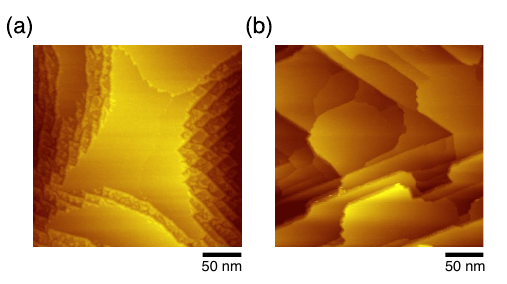}
\caption{Topography of representative regions on the Nb(100) crystal surface obtained using a Nb STM tip at (a) 1.5 K and (b) 50 mK. Image (b) shows atomically flat regions with single layer steps of about 0.3 nm. }
\label{nb}
\end{figure}

To perform the junction measurements, we used a custom STM mounted on the mixing chamber stage of a dilution refrigerator \cite{roychowdhury201430,sullivan2013asymmetric}. STM tips were made from 250 $\mu$m diameter Nb wire sharpened in an $\mathrm{SF_6}$ reactive ion etcher \cite{roychowdhury2014plasma}. Each tip was mounted on the STM and cleaned by high-voltage field emission on an Au(100) or Au(111) single crystal under vacuum at low temperature. The superconducting sample was a 4.5 mm x 4.5 mm x 1 mm Nb(100) single crystal \cite{Nb100}. The sample surface was first cleaned by Ar ion sputtering with a kinetic energy of 2 keV at a pressure of 2.0 $\times$ $10^{-5}$ mbar for 3-4 h followed by annealing at 900 $^{\circ}$C in UHV for 20 min. This cycle was repeated two or three times before the sample was transferred to the STM without breaking vacuum \cite{roychowdhury201430}. \Cref{nb} shows topographic images of the Nb surface at 1.5 K and 50 mK. Atomically flat terraces are clearly visible. 

For standard STM imaging, a voltage bias $V_b$ is applied to the sample and the tunneling current $I$ is measured. A feedback mechanism adjusts the $z$-position of the tip to maintain a constant tunneling current as it is scanned in $x$ and $y$, allowing one to image surface topography. $I(V)$ spectroscopy is obtained by setting a desired junction resistance, turning off the $z$ feedback, and sweeping $V_b$ while recording $I$. Subtle features of the $I(V)$ curve can be resolved by measuring $\didv$ directly with a lock-in amplifier at the same time. However this mode of operation does not fully characterize low loss tunnel junctions, which may be hysteretic.

\begin {figure}
\centering
\includegraphics[width=1\linewidth]{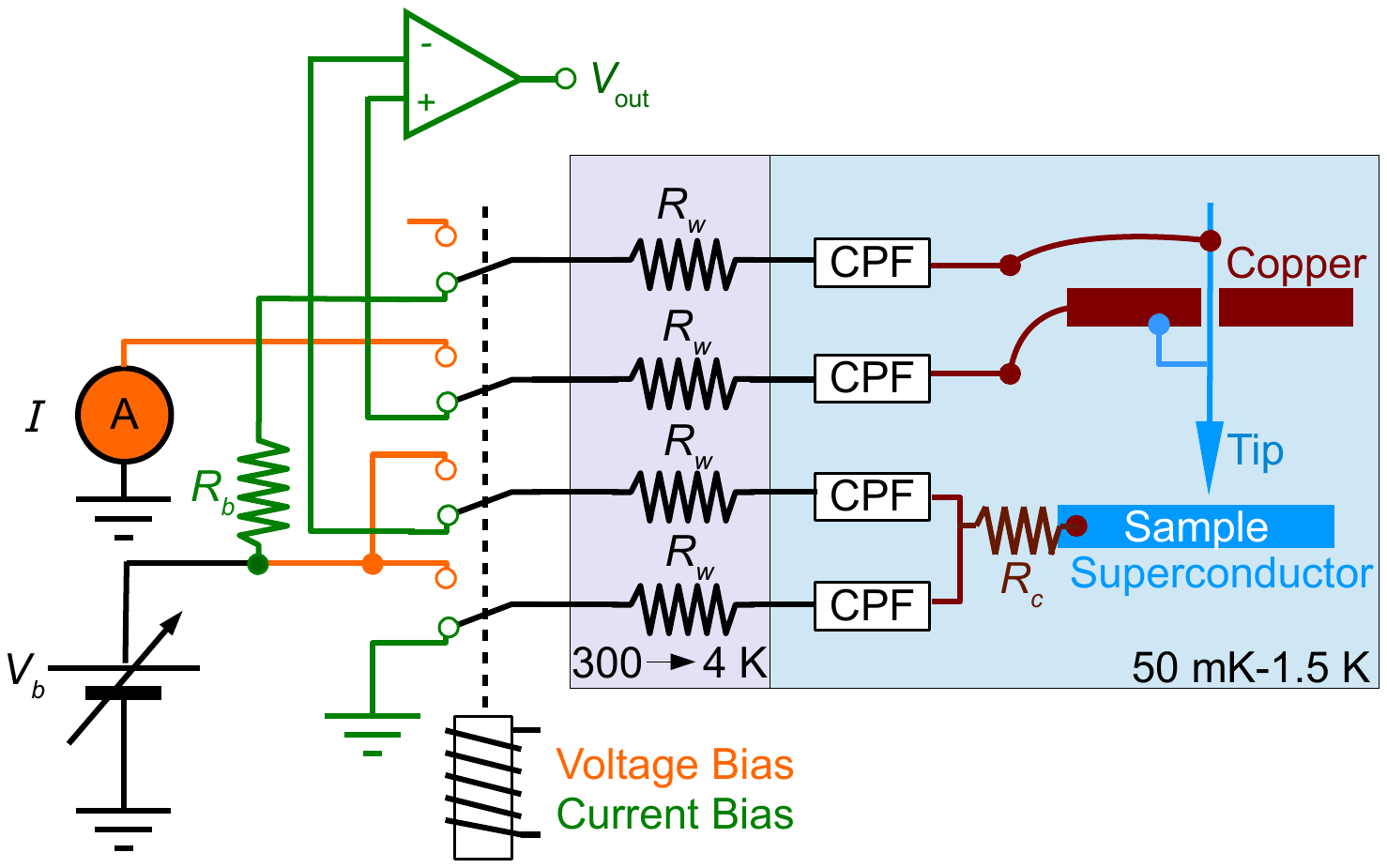}
\caption{Measurement schematic, including the relay box used to switch from voltage-biased mode to current-biased mode. The bias resistor $R_{b}$ is set to 1, 10, or 100 M$\Omega$, depending on the tunnel junction resistance. $R_{w}\approx 220$ $\Omega$ is the resistance of each line in the refrigerator and $R_{c}$ is the contact resistance between the sample mounting stud and the measurement wires. The copper powder filters (CPF) are used to filter out high frequency noise.}
\label{relay}
\end{figure}

To perform current-biased measurements, a resistor $R_b \gg R_n$ is inserted in series with the junction to effectively convert the bias voltage $V_b$ into a bias current $I_b \approx V_b/R_b$. The voltage across the junction is recorded while the bias current is swept. \Cref{relay} shows the schematic diagram of the voltage- and current-biased circuits we used, with a relay that switches between the two while the tip feedback is turned off. To clearly distinguish between these two types of measurements, we specifically use $I(V)$ for voltage-biased curves (orange path in \cref{relay}), and $V(I)$ for current-biased curves (green path).

\section{\label{sec:level4}Current-Voltage characteristics} 
We next describe measurements of the $I-V$ characteristics for a range of the coupling between the tip and sample. To set each junction, feedback was established at a target setpoint current with a fixed voltage bias $V_b = 3.5\ \textrm{mV}$. The feedback was then turned off and we measured voltage-biased $I(V)$ and $\didv$. We then switched to the current-biased mode to measure $V(I)$ for the same junction. Both measurements were taken with forward (\textit{i.e.} increasing the current from negative to positive) and reverse sweeps. The curves were corrected to account for a contact resistance $R_c$ (typically under $100\ \Omega$) between the sample stud and the measurement wires. We refer to each set of curves by the normal-state resistance $R_n$ extracted from the characteristics at high voltage, where they are nearly ohmic.

\begin {figure*}[t]
\centering
\includegraphics[width=\textwidth]{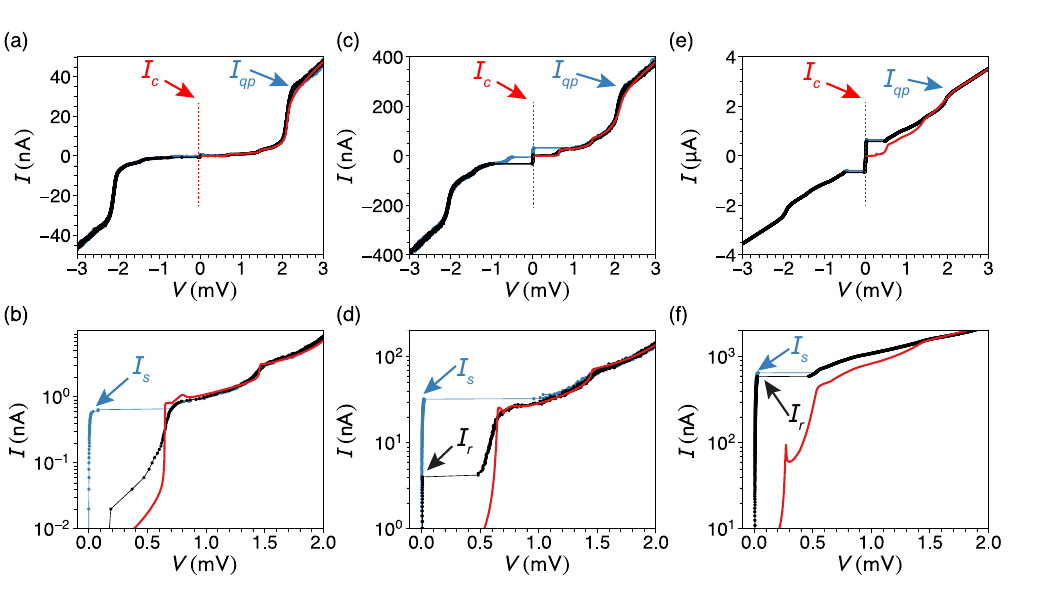}
\caption{$V(I)$ characteristics for junction tunneling resistance $R_n$ of (a) and (b) 57 k$\Omega$, (c) and (d) 7.9 k$\Omega$, and (e) and (f) 1.08 k$\Omega$. Linear plots are shown in the first row, while the second row shows semilog graphs, revealing details of the switching and the retrapping behaviors. Each panel shows forward (blue) and reverse (black) sweeps measured at 1.5 K, along with a fit curve from the asymmetric MAR theory of \cref{eq:mIN} (red solid line). Also shown is the ideal supercurrent branch at zero voltage (red dashed line), which ends at the critical current $I_c$, predicted from MAR theory. See \cref{tab} for fit parameters. }
\label{iv}
\end{figure*}

\Cref{iv} shows $V(I)$ characteristics for three values of $R_n$, acquired at 1.5 K, with corresponding $I(V)$ and \didv\ curves shown in the Supplemental Material \cite{ref1}. \Cref{iv}(a) was acquired at $R_n = 57\ \textrm{k}\Omega$ and shows conventional low-transparency S-I-S behavior.  As the current is reduced from its maximum value in the reverse sweep, the characteristic is roughly linear, with a slope of $1/R_n$.  At $\sim 2.1\ \textrm{mV}$, corresponding to the sum of the superconducting gaps, the current drops to near zero. We identify the quasiparticle current $I_{qp} = 33$ nA as the current at the intersection of the ohmic section of the voltage branch and the sudden rise at the gap edge. The current in the subgap region is small, with steps around 0.67 mV and 1.50 mV as seen more clearly in \cref{iv}(b), indicating multiple Andreev reflection behavior. When the current increases from zero in the forward sweep, the junction starts in the supercurrent state with a small resistance and abruptly jumps to the voltage branch at a switching current $I_{s} = 620\ \textrm{pA}$.

In order to model the measured $I-V$ curves with the asymmetric MAR theory, we assume a total current $I(V) = \displaystyle\sum_{i=1}^N I_0(D_i, \Delta_{s}, \Delta_{t}, T, V)$ carried by $N$ independent conduction channels. Here, $I_0$ is the dc current calculated from Eq.~\eqref{eq:sis:cthe4}, each channel has a distinct transparency $D_i$, and the energy gaps of the STM sample and tip are $\Delta_{s}$ and $\Delta_{t}$. In this case, the total conductance is given by the Landauer formula as $G = 1/{R_{n}} = G_0\displaystyle\sum_{i=1}^{N} D_i$. Whereas previous analyses decomposed the total conductance into channels with individually fitted transparencies \cite{scheer1997conduction, scheer1998signature, scheer2001proximity, hiraoka2014transport, ludoph2000multiple}, we find that our measurements can be described by assuming a single transparency shared across all channels, with an example shown in the Supplemental Material \cite{ref1}. Thus a single transparency value can be used to characterize the transport at each $R_n$. The current can then be simplified as
\begin{align}
I(V) = N I_0(D, \Delta_{s}, \Delta_{t}, T, V),
\label{eq:mIN}
\end{align}
where $N$ specifies a non-integral number of effective channels. The corresponding normal state resistance is $R_n = 1 / (G_0 D N)$.
 
We chose values of $\Delta_{s}$, $\Delta_{t}$, $D$, and $N$ to reproduce the location of peaks in \didv, the level of the subgap current in $V(I)$, and the resistance of the voltage branch, with $T$ fixed at the nominal refrigerator temperature. The parameter values are listed in \cref{tab}, with the MAR characteristic calculated from \cref{eq:mIN} shown in \cref{iv}. For the $R_n$ = 57 k$\Omega$ data, we found that the step-like subgap features and quasiparticle current rise were reproduced for $D$ = 0.1 (and thus $N$ = 2.26) and superconducting gaps of 1.497 and 0.67 meV. The smaller gap is likely that of the tip, because the etching process can alter its superconducting properties. A tip gap of 0.67 meV suggests a superconducting transition temperature $T_{c} \approx 4.4$ K, consistent with 5$\%$ oxygen in Nb \cite{PhysRevB.9.888}, which is below the solubility limit of 9$\%$. 

With the four MAR fitting parameters, \cref{ic} predicts a critical current $I_c$ = 26 nA, as indicated in \cref{iv}(a). The observed switching current $I_{s}=$ 620 pA is two orders of magnitude smaller. Such large suppression has also been reported in atomic scale Nb-Nb mechanically controllable break junctions \cite{ludoph2000multiple} and recent current-biasd STM measurements \cite{jack2017quantum,trahms2023diode}. Note here the switching and retrapping currents are not true supercurrents because this branch of the $V(I)$ curve has a non-zero resistance of $R_{z} \approx 54.8$ k$\Omega$, which will be discussed at the end of the next section. 

\Cref{iv}(c) was taken at a junction resistance of 7.9 k$\Omega$. The lower $R_n$ was set by using a larger feedback current, corresponding to a smaller tip-sample separation. As seen in Table I, while $\Delta_s$ only drops slightly compared to $R_n$ = 57 k$\Omega$, $\Delta_t$ decreases significantly. $D$ increases from 0.1 to 0.27, resulting in $N=6.05$. The subgap currents are much larger than \cref{iv}(a), as expected from $D$ being larger, representing a more transparent tunnel barrier. The switching current in \cref{iv}(d) is $I_{s} \approx 32$ nA, while the calculated $I_{c} \approx 224$ nA. Notice that the switching current in \cref{iv}(c) is about 50 times larger than the switching current in \cref{iv}(a), even though $R_{n}$ differs by only a factor of 7. \Cref{iv}(d) also shows hysteretic behavior with the voltage abruptly dropping to zero during the reverse sweep at a retrapping current $I_{r}$ of 4.6 nA.

\Cref{iv}(e) shows data for $R_n=1.08$ k$\Omega$, where the tip-sample distance was reduced further. The characteristic shows an even larger subgap current, where $I_{qp}$ has increased to nearly 2.6 $\mu$A, while the switching current has increased to 660 nA. In this regime, unlike the highly hysteretic behavior in \cref{iv}(c), $I_{r}=595$ nA is about 90$\%$ of the switching current $I_{s}$, showing a higher ratio of retrapping current with respect to the switching current for larger damping \cite{Orlando}. This phenomena is likely due to the tip and the sample being in point contact (Supplemental Material \cite{ref1}). The fit curve in \cref{iv}(e) disagrees noticeably with the data below 1.5 mV, and is very sensitive to and the transparency parameter $D$. As one can see, the simplified multichannel fit does not exactly reproduce the Andreev steps inside the gap. In this case, it would be natural to expect multiple channels with different transparencies $D_i$ for low junction resistance \cite{scheer2001proximity,ludoph2000multiple,hiraoka2014transport,scheer1998signature,scheer1997conduction}.

We took similar measurements for $R_n$ between 1 k$\Omega$ and 10 M$\Omega$.  We begin the discussion of the extracted junction properties with the characteristic currents shown in \cref{main}. Here, $I_{qp}$, $I_s$, and $I_r$ were found directly from the $V(I)$ curves and $I_c$ was calculated from the asymmetric MAR theory using the parameters $\Delta_s$, $\Delta_t$, $D$, $N$ for each curve. We acquired data at 50 mK and 1.5 K to check for thermal effects and found generally good agreement between the two temperatures.

 For $R_n$ $>$ 100 k$\Omega$, no switching or retrapping currents were observed, consistent with thermal fluctuations destroying phase coherence \cite{proslier2006probing,PhysRevB.78.140507,PhysRevLett.101.037002}. For 10 k$\Omega<$ $R_n$ $<$ 100 k$\Omega$, a small supercurrent branch and hysteretic behavior were observed. In this regime, the switching current $I_{s}$ ranges from 1$\%$ to 10$\%$ of the predicted critical current $I_{c}$ (red circles), and increases rapidly as the junction resistance $R_n$ decreases, scaling approximately as $R_n^{-2}$. The retrapping current $I_{r}$ (triangles) increases much more rapidly than the switching current $I_{s}$ as $R_n$ decreases. Similar behavior has been observed in Refs.~\cite{jack2017quantum, lu2023phase}. For $R_n$ $<$ 3 k$\Omega$, where we will argue that the tip and the sample have made solid contact, $I_{r}$ $\approx$ $I_{s}$. In this same range of $R_n$, there is a clear disagreement between the MAR theory and experimental curve in the subgap regime below 1.5 mV.

To precisely identify the superconducting gaps of the tip and sample, we analyzed the peaks in the $\didv$ data. While MAR results in a series of steps in the $I-V$ curves, and thus peaks in $\didv$, the largest occur near voltages corresponding to energies $\Delta_t$, $\Delta_s$, and $\Delta_s + \Delta_t$.  We refer to these peak locations as the voltages $V_t$, $V_s$, and $V_{st}$. In \cref{dIdVth}, the extracted values for these three voltages are shown as a function of $D$, where the transparency values were obtained by fitting the MAR theory to the corresponding $V(I)$ curves. The peak voltages decrease with increasing $D$ (i.e. decreasing $R_n$), particularly for $V_{st}$.

The figure also shows theoretical values of $\didv$ calculated from \cref{eq:mIN} with fixed gap values of  $\Delta_t = 0.657$ meV and $\Delta_s = 1.514$ meV (as indicated by the dashed lines). The peaks in $\didv$ (solid lines) occur at the gap energies in the $D = 0$ tunnel limit, but shift to lower voltages with increasing transparency due to the presence of increasingly large subgap currents. A similar effect was reported in Ref. \cite{PhysRevB.74.132501}. The predicted shift, however, is smaller than what is seen in the data, suggesting that the gaps themselves were changing with $D$ in our STM measurements.

\Cref{DR}(a) shows the same peak values, though now plotted as a function of $R_n$. Also shown are the values of $\Delta_t$ and $\Delta_s$ needed for the MAR theory to best reproduce the measured voltages $V_t$, $V_s$, and $V_{st}$ (with $D$ fixed at the fit values), where the agreement is to within 10 $\mu$V. $\Delta_s$ is roughly constant, with a value near that of bulk Nb, while $\Delta_t$ decreases by almost 15\% as $R_n$ decreases. It shows that while some of the reduction in $V_t$ might be coming from the bending of the MAR theory prediction, the actual gap of the tip is changing.  

There are several reasons why the tip gap could decrease below $R_n=$ 5 k$\Omega$, where the junction transparency is large. One possible explanation is that the tip starts to make contact with the sample, and the tip compresses more than the sample due to its geometry. This applies pressure to the material at the end of the tip, reducing its $T_{c}$. It might also be due to the fact that pure Nb from the sample is less sensitive to pressure than oxygen-doped Nb from the tip. A third possible explanation is that a high current density in the tip causes heating, since the power dissipation is roughly 100 times higher at low $D$ than high $D$. The small tip not only has a smaller thermal conductance than the sample, but if it also has a smaller $T_{c}$ due to contaminants, this will lead to an even larger reduction in the superconducting gap from heating. 

\Cref{DR}(b) shows the extracted transparency $D$ and effective number of conduction channels $N$ as a function of $R_n$ from the data set in \cref{main}. As expected, $D \ll 1$ when the junction resistance $R_n$ is high, indicating very low transparency and a junction in the standard S-I-S tunneling limit \cite{tinkham12}. As the tip and sample get closer, $R_n$ decreases and the transparency $D$ becomes larger. Interestingly, $D$ does not reach 1 for low $R_n$ but saturates at around 0.6. The effective number of channels never decreases below 2. Similar phenomena were seen in a previous study of niobium \cite{scheer1998signature}. On the other hand, when the junction resistance was small, we clearly see that $N$ increases, as does the transparency $D$, indicating a higher overall transmission of the junction. 

Also shown in \Cref{DR}(b) is the transparency $D$ = 1 /$(G_0 R_n N)$ as a function of $R_n$ for fixed values of $N$. For $N = 1$, this shows the maximum transparency that any channel can have for a given value of $R_n$. We were not able to extract $D$ and $N$ for $R_n>$ 70 k$\Omega$ because the subgap currents in $V(I)$ were too small to fit the MAR theory and obtain accurate results.

\section{\label{sec:level6}Underdamped junction properties}
Our observation of a relatively large retrapping current $I_r$ over a large range of $R_n$ indicates that the junction was subject to a certain amount of dissipation. Here we compare our results to the standard resistively and capacitively shunted Josephson junction (RCSJ) model \cite{tinkham12} to extract additional junction parameters.

In the RCSJ model, the superconducting phase difference across an unbiased junction will oscillate at the plasma frequency 
\begin{align}
\omega_{p}=\sqrt{2 \pi I_c/{\Phi_0 C}}
\label{equation:omega}
\end{align}
with a quality factor $Q=\omega_{p}R_eC=\sqrt{\beta_{c}}$. Here, $\beta_{c}=2\pi I_{c}R_{e}^2C/\Phi_0$ is the Stewart-McCumber hysteresis parameter, $R_{e}$ is the total effective resistance shunting the Josephson junction, $C$ is the total effective capacitance across the junction, and $\Phi_0=h/2e$ is the magnetic flux quantum. If the damping is small, with $\beta_{c}>1$, the junction is said to be underdamped.

\begin {figure}[t]
\centering
\includegraphics[width=1.0\linewidth]{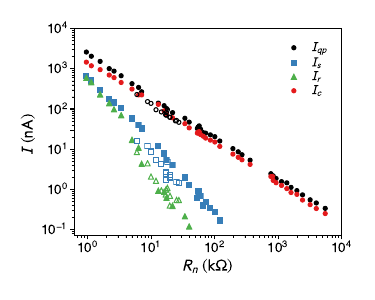}
\caption{Extracted quasiparticle current $I_{qp}$ (black circles), switching current $I_{s}$ (blue squares), and retrapping current $I_{r}$ (green triangles) as a function of $R_n$, for data acquired at 50 mK (open symbols) and 1.5 K (filled). The theoretical critical current $I_c$ (red) was calculated from the asymmetric MAR theory using the fitted parameters $R_n$, $\Delta_s$, $\Delta_t$, and $D$ for each curve. For $R_n > 105$ k$\Omega$, there was no observed supercurrent branch, so $I_s$ and $I_r$ could not be determined. For $R_n <$ 3 k$\Omega$, hysteretic behavior disappeared. }
\label{main}
\end{figure}

\begin {figure}[t]
\centering
\includegraphics[width=1.\linewidth]{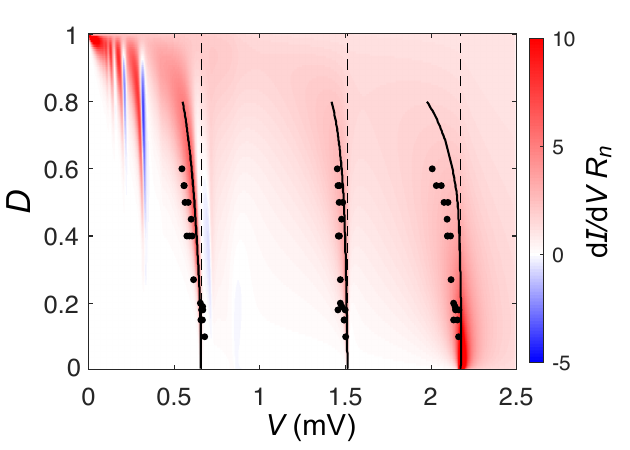}
\caption{$\didv$ peak locations $V_t$, $V_s$, and $V_{st}$ (black circles) extracted from the data, on top of theoretical values of $\didv$ (pseudocolor plot), normalized by $R_n$, for $\Delta_t=0.657$ meV and $\Delta_s=1.514$ meV at $T=1.5$ K. The peaks extracted from the theory (solid curve) deviate from the fixed gaps (dashed) with increasing transparency.}
\label{dIdVth}
\end{figure}

\begin {figure}[t]
\centering
\includegraphics[width=\linewidth,height=0.75\textheight,keepaspectratio]{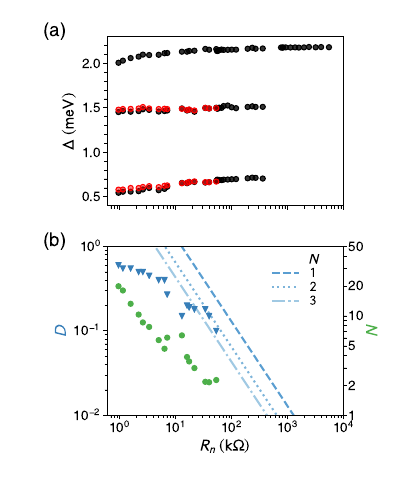}
\caption{(a) Black circles are the peaks $eV_s$, $eV_t$, and $eV_{st}$, as a function of $R_n$, extracted from the experimental $\didv$ data. Red circles show $\Delta_{s}$, $\Delta_{t}$ as a function of $R_n$, fitted from the asymmetric MAR theory by minimizing the difference between the theoretical $\didv$ peaks and the measured $\didv$ peaks. (b) Log-log plot of the extracted transparency $D$ (blue triangles) and effective number of channels $N$ (green circles) as a function of $R_n$, from fitting the asymmetric MAR theory to $V(I)$ curves. The dashed lines are the theoretical transparency $D$ as a function of junction resistance $R_n$ for fixed values of $N$.}
\label{DR}
\end{figure}

\begin{figure}[t]
\centering
\includegraphics[width=\linewidth,height=0.75\textheight,keepaspectratio]{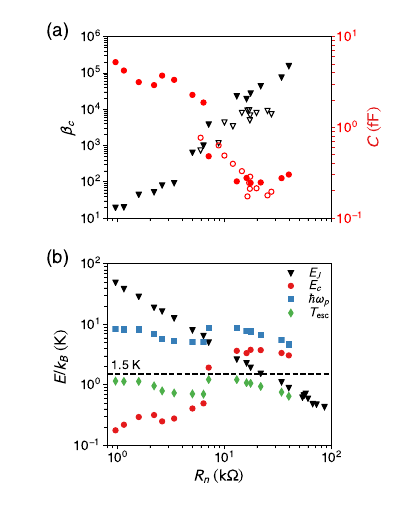}
\caption{(a) Extracted values of the Stewart-McCumber parameter $\beta_{c}$ (black triangles) and the total effective capacitance $C$ (red circles) of the junction versus tunneling resistance $R_n$, for data acquired at 50 mK (open symbols) and 1.5 K (filled). (b) Josephson energy $E_{J}$ (black triangles), charging energy $E_{c}$ (red circles) and plasma energy $\hbar \omega_{p}$ (blue squares) versus normal resistance $R_n$ of the junction. The green diamonds show the MQT escape temperature $T_\textrm{esc}$. The higher temperature at which data were acquired is shown for reference (dashed line). }
\label{scale}
\end{figure}

In terms of the retrapping current $I_r$, the damping parameter is \cite{chen1988return,tinkham12}
\begin{align}
\beta_{c}=(I_{qp}/I_{r})^2.
\label{equation:betac}
\end{align}
\Cref{scale}(a) shows values of $\beta_{c}$ (black triangles) as a function of $R_n$ at 1.5 K, extracted using this expression. Since $\beta_{c}\gg1 $ for all $R_n$, our junctions are in the relatively low dissipation limit. Using these values of $\beta_{c}$ and assuming $I_{c}=\pi I_{qp}/4$, the junction capacitance is
\begin{equation}
 C=4\beta_{c}\hbar/2eI_{qp}R_e^2\pi,
 \end{equation}
where $I_{qp}$ is identified directly from the data (black circles in \cref{main}), and the effective resistance at the retrapping voltage $V_{r}$ is
\begin{equation}
R_{e} = V_{r}/I_{r}.
 \end{equation}

\Cref{scale}(a) also shows the resulting capacitance $C$ as a function of $R_n$ (red circles). The capacitance is in reasonable agreement with what one would expect for an STM tip that is near a conducting surface \cite{kurokawa1997tip,de2017fast,kurokawa1998gap, PhysRevB.93.020504}; the values are on the order of fF and increase as the separation between the tip and sample becomes smaller. For $R_n \geq 45$ k$\Omega$, we could not use this approach to determine $C$ because the retrapping current became too small to measure.

Notice that all our curves in \cref{main} show early switching out of the supercurrent state. While the RCSJ model with MAR explains much of the observed behavior along the voltage branch, small Josephson junction physics is needed to explain the supercurrent branch. The dynamics of Josephson junctions are governed by the level of dissipation and the relative size of four different energies: the Coulomb charging energy $E_{c}= e^2/2C$, the Josephson energy $E_{J}=I_{c}\Phi_0/2\pi$, the thermal energy $k_{B}T$, and the plasma energy $\hbar\omega_{p}$.

\Cref{scale}(b) shows these energies as a function of the tunneling resistance $R_n$. For $R_n<10$ k$\Omega$, $E_{J}>\hbar\omega_{p}>E_{c}$. In this limit, even though the quantum mechanical phase difference across the junction is relatively well defined, macroscopic quantum tunneling (MQT) \cite{tinkham12, PhysRevB.39.6465, PhysRevLett.55.1908, PhysRevB.35.4682} can produce early switching of the supercurrent. Although there was a strong dependence of the junction behavior on $R_n$, we see little difference between the data taken at 50 mK (open symbols) and 1.5 K (filled symbols) in \cref{main}. The thermal energy of the junction at 50 mK was nominally 30 times smaller than at 1.5 K, but $I_{s}$ was not significantly larger at 50 mK, nor was $I_{r}$ significantly smaller at 50 mK.

This suggests that the switching current in our STM junction was not limited by equilibrium thermal fluctuations. One possibility is MQT of the phase, due to the small critical current and capacitance of the junction. Another possibility is external noise driving the system to switch into the voltage state. A third possibility is that the junction is at an effective temperature that is well above the refrigerator temperature due to heating in the leads from the applied bias current. Although we can not rule out the last two alternatives, our measurements show relatively low noise.

If the junction was well-isolated from dissipation (including radiation loss), then MQT would yield an effective switching temperature of $T_\textrm{esc}=\hbar\omega_{p}/7.2k_{B}$ \cite{PhysRevLett.55.1908,PhysRevB.35.4682}. As \cref{scale}(b) shows, $T_\textrm{esc}\approx 1$ K for a wide range of $R_n$, which suggests MQT may be important and producing the switching currents at 50 mK that are comparable to those at 1.5 K (see \cref{main}); both are much less than the critical current obtained by fitting the MAR theory to the measured $V(I)$ curves. For $R_n$ greater than about 10 k$\Omega$, $E_{c}$ dominates $E_{J}$. In this limit, charging effects, phase diffusion and the impedance of the bias leads become important in determining the $V(I)$ characteristic, including in particular the size of the observed phase-coherent supercurrent and the resistance of the supercurrent branch. 

For small Josephson junctions, it is important to consider the effects produced by the impedance of the leads \cite{anchenko1969josephson}. The plasma frequency $\omega_{p}$ of a small junction is typically $10^{10}$ Hz to $10^{11}$ Hz and at such frequencies the shunting impedance $Z_1$ of typical bias leads has non-zero real and imaginary parts that are of order 50 $\Omega$, which is orders of magnitude smaller than the tunneling resistance we observed. The output leads act like a mismatched transmission line, leading to high-frequency damping that increases the retrapping current and creates a non-zero resistance on the supercurrent branch due to the inelastic tunneling of Cooper pairs \cite{PhysRevB.39.6465,PhysRevB.35.4682,tinkham12,PhysRevLett.53.1260, lu2023phase}. High frequency damping also produces noise, which causes phase diffusion, and an average voltage on the supercurrent branch. For thermally generated Nyquist noise and $E_{J} \ll k_{B}T$, one finds the supercurrent branch has a resistance \cite{PhysRevB.50.395,tinkham12}
\begin{equation}
R_{z}=2Z_1(k_{B}T/E_{J})^2=2Z_1\bigg[\frac{4e^2k_{B}T\Rn}{\hbar \pi \Delta}\bigg]^2.
\label{eq:r0}
\end{equation}

\begin{figure}[t]
\centering
\includegraphics[width=1.0\linewidth]{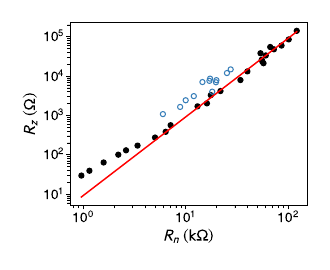}
\caption{Resistance of the supercurrent branch $R_z$ versus tunnel resistance $R_n$ of the junction. The values extracted from measurements at 50 mK (open blue circles) and 1.5 K (black circles) roughly scale as $R_z \propto R_n^2$ (red).}
\label{r0} 
\end{figure}

\Cref{r0} shows a plot of $R_{z}$ as a function of $R_n$, where $R_{z}$ is the supercurrent branch resistance after subtracting off the contact resistance $R_{c}$ between the sample stud and bias lines. We estimated $R_{c} \approx 173$ $\Omega$ at 1.5 K and $R_{c}\approx 0$ $\Omega$ at 50 mK. The plot shows that $R_{z}$ scales quadratically with $R_n$, consistent with \cref{eq:r0}, suggesting that the critical current is being suppressed by phase diffusion. The extracted value of $Z_1$ is 820 at 1.5 K and 1776 at 50 mK.

\section{\label{sec7}CONCLUSION}
In conclusion, we used a Nb tip and Nb sample in a low temperature STM to measure the current-voltage characteristics of small superconducting Josephson junctions. By adjusting the distance between the tip and the sample, we varied the junction tunneling resistance $R_n$ and observed $I-V$ characteristics that ranged from the low-transparency S-I-S tunneling limit to the high-transparency limit, and the superconducting gap of the tip decreased by 15$\%$ in this limit. We compared our results to simulations from a generalized Averin-Bardas model of MAR in junctions with different superconducting energy gaps. We find that the asymmetrical MAR junction theory is important for reliably extracting the superconducting gaps and the transparency of the junctions. 

For $R_n$ between 10 k$\Omega$ and 100 k$\Omega$, the junction behaved like a hysteretic small Josephson junction with a suppressed switching current and a small but non-zero resistance $R_z$ on the supercurrent branch. We note, in principle the full supercurrent could be observed by stabilizing the phase using a SQUID-STM configuration \cite{PhysRevApplied.4.034011,sullivan2013asymmetric,liao2017simultaneously,van1997probing,VANHARLINGEN1999410}. For $R_n < 10$ k$\Omega$, the junction acted like a point contact. Although our MAR model does a good job of explaining many features of the $I-V$ characteristics, it leaves out potentially important physical effects, including phase diffusion, charging effects, noise, incoherent pair tunneling, and MQT. A comprehensive theory that includes all of this relevant physics, as well as MAR, is currently not available but is essential for interpreting superconducting STM measurements and characteristics of other superconducting devices with small tunnel junctions, such as qubits \cite{ridderbos2019multiple, PhysRevLett.121.047001, PhysRevX.9.011010,lu2023phase}. 

\section{\label{sec8}Acknowledgments}
We thank D. Averin for fruitful discussions and providing the MAR code for the symmetric junction, as well as Z. Steffen and B. S. Palmer for offering useful insights on the MAR calculations. We acknowledge support from the Laboratory for Physical Sciences, the JQI, CNAM, and from the NSF for supporting development of the STM under grant DMR-1409925.

\bibliography{aBibtex2}
\bibliographystyle{apsrev4-2}

\clearpage
\newpage

\begin{table}
\caption{Extracted parameters of the normal state tunneling resistance $R_n$, superconducting energy gaps $\Delta_{s}$ and $\Delta_{t}$, transparency $D$, and effective number of channels $N$, for the $V(I)$ characteristics in \cref{iv}}
 \centering
\begin{ruledtabular}
\begin{tabular}{lccccccc}
$R_n$ (k$\Omega$) & $\Delta_{s}$ (meV) & $\Delta_{t}$ (meV) &  $D$  & $N$\\\hline
57  & 1.497 & 0.670& 0.1   & 2.26\\
7.9  & 1.486 & 0.627 & 0.27    & 6.05\\  
1.08  & 1.479 & 0.579 & 0.6   & 19.91     
\end{tabular}
\end{ruledtabular}
\label{tab}
\end{table}

\clearpage
\appendix
\section*{Supplemental Material}

\setcounter{figure}{0}
\renewcommand{\thefigure}{S\arabic{figure}}
\renewcommand{\theHfigure}{S\arabic{figure}} 

\setcounter{table}{0}
\renewcommand{\thetable}{S\arabic{table}}
\renewcommand{\theHtable}{S\arabic{table}}

\setcounter{equation}{0}
\renewcommand{\theequation}{S\arabic{equation}}
\renewcommand{\theHequation}{S\arabic{equation}}

\renewcommand\thefigure{\textrm{S.}\arabic{figure}}

\title{Multiple Andreev Reflection Effects in Asymmetric STM Josephson Junctions}

\author{Wan-Ting Liao}
\affiliation{Laboratory for Physical Sciences, College Park, Maryland 20740}
\affiliation{Quantum Materials Center, Department of Physics, University of Maryland, College Park, Maryland 20742}
\author{S. K. Dutta}
\affiliation{Quantum Materials Center, Department of Physics, University of Maryland, College Park, Maryland 20742}
\author{R. E. Butera}
\affiliation{Laboratory for Physical Sciences, College Park, Maryland 20740}
\author{C. J. Lobb}
\affiliation{Quantum Materials Center, Department of Physics, University of Maryland, College Park, Maryland 20742}
\affiliation{Joint Quantum Institute, University of Maryland, College Park, Maryland 20742}
\author{F. C. Wellstood}
\affiliation{Quantum Materials Center, Department of Physics, University of Maryland, College Park, Maryland 20742}
\affiliation{Joint Quantum Institute, University of Maryland, College Park, Maryland 20742}
\author{M. Dreyer}
\affiliation{Laboratory for Physical Sciences, College Park, Maryland 20740}
\date{\today}

\maketitle

\section{Multiple Andreev Reflection}
In Fig.~1 of the main text, the schematic drawing of the Andreev process only shows the wavefunction components for the left and the right superconductor. Here, we show example trajectories for an electron-like source quasiparticle incident from the left superconductor. Note that the superconducting gap is different between the left and the right superconductors. The voltage bias $V$ is applied to the right superconducting electrode $S_R$ with respect to the left superconducting electrode $S_L$. 

An electron-like source quasiparticle incident from the left superconductor with energy $\varepsilon$ will travel to the right interface barrier due to the applied voltage $V$, it goes through the center barrier, with partial backward reflection, and partial forward transmission. When it reaches the right superconductor, the electron-like quasiparticle has picked up energy $eV$, resulting in an energy $(\varepsilon+eV)$. At the right interface $N_R-S_R$, it goes through the Andreev reflection process with amplitude $a^R(\varepsilon+eV)$ and emerges as a hole-like quasiparticle with energy $-(\varepsilon+eV)$.

The hole-like quasiparticle will then travel towards the left electrode. It encounters the barrier again, resulting in partial transmission and partial reflection. The transmitted hole-like quasiparticle travels back to the left interface $N_L-S_L$, picking up extra energy $eV$, becoming $-(\varepsilon+2eV)$. At the interface $N_L-S_L$, the hole-like quasiparticle is again Andreev reflected back as electron-like quasiparticle, with energy $\varepsilon+2eV$. This process continues, allowing the iterative construction of the electron and hole wavefunctions in the normal regions. 

\begin {figure}[t]
\centering
\includegraphics[width=0.9\linewidth]{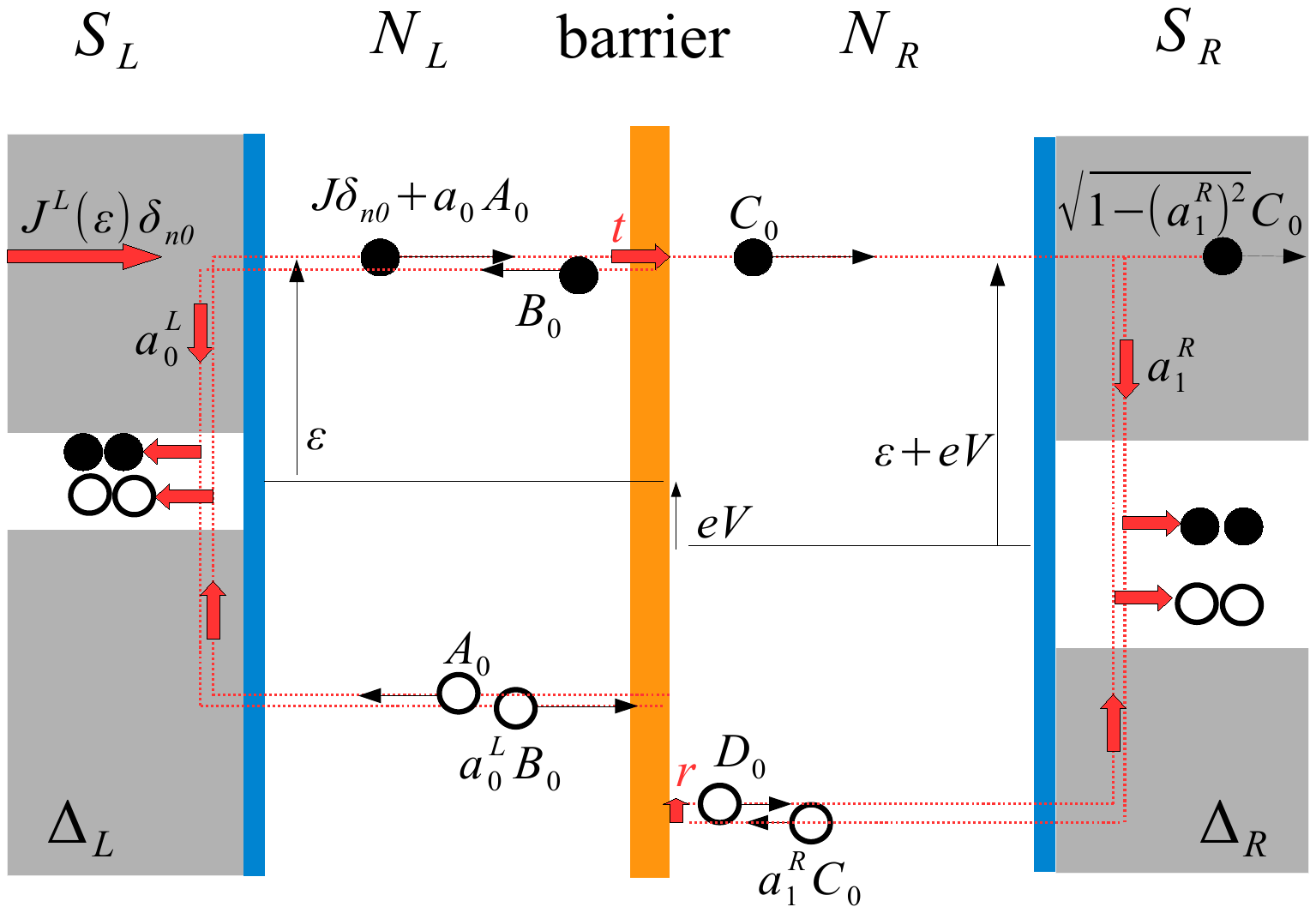}
\caption{Illustration showing MAR in the generalized AB model for the case of electron-like source quasiparticles incident from the left. The barrier (orange) in the center acts as a scattering region for quasiparticles and is characterized by transparency $D$ (distinct from the incident hole term $D_n$). The superconductor-to-normal metal interfaces (blue) have unity transparency. The wavefunction amplitudes $C_n$, and $D_n$ at the right interface are shown. The superscript in $A_{n}$, $B_{n}$, $C_{n}$, $D_{n}$ is not labeled here for ease of visualizing the complex process.}
\label{averin2}
\end{figure}

\clearpage
\section{MAR wavefunctions for a quasiparticle incident from the right}

In Eqs.~(3) and (4) of the main text, we listed the final electron and hole wavefunctions for an electron-like quasiparticle incident from the left and a hole-like quasiparticle incident from the left. Here we write down the wavefunctions for an electron-like quasiparticle incident from the right, and a hole-like quasiparticle incident from the right. For consistency, $k$ is the Fermi wavevector associated with the quasiparticle at energy $\varepsilon$, positive by definition. Thus an electron traveling to the right has the same value of $k$ as a hole traveling to the left.  

The wavefunctions in the normal region for an electron-like quasiparticle incident from the right when $D \neq 1$ are
\begin{widetext}
\begin{align}
   \psi_{e}^{eR}&=\displaystyle\sum_{n=-\infty}^{\infty} [(a_{\{2n,- \}}^{R} A_n^{\leftarrow}+J^{R} \mathit{\delta}_{n0})e^{-ikx}+B_n^{\leftarrow}e^{+ikx}]e^{-i(\varepsilon-2neV)t/\hbar}\nonumber\\
    \psi_{h}^{eR}&=\displaystyle\sum_{n=-\infty}^{\infty} [A_n^{\leftarrow}e^{-ikx}+a_{\{2n,- \}}^{R}B_n^{\leftarrow} e^{+ ikx}]e^{-i(\varepsilon-2neV)t/\hbar}, 
\label{sup:er}
\end{align}
\end{widetext}
where $a_{\{2n,- \}}^{R}=a^{R}(\varepsilon-2neV)$ and $J^{R}(\varepsilon)=\sqrt{[1-|a^{R}(\varepsilon)|^2]}$.

The wavefunctions in the normal region for a hole-like quasiparticle incident from the right are
\begin{widetext}
\begin{align}
   \psi_{h}^{hR}&=\displaystyle\sum_{n=-\infty}^{\infty} [(a_{\{2n,+ \}}^{R} A_n^{\leftarrow}+J^{R} \mathit{\delta}_{n0})e^{+ ikx}+B_n^{\leftarrow}e^{- ikx}]e^{ -i(\varepsilon+2neV)t/\hbar}\nonumber\\
    \psi_{e}^{hR}&=\displaystyle\sum_{n=-\infty}^{\infty} [A_n^{\leftarrow}e^{+ikx}+a_{\{2n,+ \}}^{R}B_n^{\leftarrow} e^{- ikx}]e^{-i(\varepsilon+2neV)t/\hbar}, 
\label{sup:hr}
\end{align} 
\end{widetext}
where $a_{\{2n,+ \}}^{R}=a^{R}(\varepsilon+2neV)$. For electron-like and hole-like source quasiparticles incident from the right, the amplitudes are labeled as $A_n^{\leftarrow}$ and $B_n^{\leftarrow}$. 

The total probability current density is then 
\begin{widetext}
\begin{align}
j_{R}
&=\frac{\hbar}{2mi}[(\psi_{e}^{eR*} \nabla\psi_{e}^{eR}-\psi_{e}^{eR}\nabla \psi_{e}^{eR*})+(\psi_{h}^{eR*} \nabla\psi_{h}^{eR}-\psi_{h}^{eR}\nabla \psi_{h}^{eR*})] \nonumber\\
&+\frac{\hbar}{2mi}[(\psi_{e}^{hR*} \nabla\psi_{e}^{hR}-\psi_{e}^{hR}\nabla \psi_{e}^{hR*})+(\psi_{h}^{hR*} \nabla\psi_{h}^{hR}-\psi_{h}^{hR}\nabla \psi_{h}^{hR*})].
    \label{eq:sis:wvd2_s}
\end{align}
\end{widetext}

The total electrical current due to an electron and hole incident from the right is directly proportional to the electron probability current density. Summing over the source energies,
\begin{align}
I_{R}(t)&=\frac{e}{2\pi \hbar}\frac{m}{\hbar k} \int_{-\infty}^{\infty}d\varepsilon  j_{R}(\varepsilon,t) f(\varepsilon) ,
\label{}
\end{align}
where $f(\varepsilon$)=$(e^{\varepsilon/k_BT}+1)^{-1}$ is the electron occupancy for states at energy $\varepsilon$, which we assume is a Fermi distribution.

\section{Recursion Relation derivation}

The recursion relations are introduced solely to eliminate the degrees of freedom of the left normal region by connecting the right normal region through the normal-state scattering matrix; as a result, their form is independent of the asymmetry of the superconducting gaps and remains identical to the symmetric-gap case. For simplicity, we will assume the superconducting gaps of the left and right electrode are the same in this section, and will omit the $\rightarrow{}$ in the wavefunctions. 

The wavefunctions in region $N_{L}$ produced by an electron-like quasiparticle incident from the left can be written as \cite{averin1995ac} 
\begin{align}
   \psi_{e}&=\displaystyle\sum_{n} [(a_{2n} A_n+J \delta_{n0})e^{ikx}+B_ne^{-ikx}]e^{-i(\varepsilon+2neV)t/\hbar}\nonumber\\
   \psi_{h}&=\displaystyle\sum_{n} [A_ne^{ikx}+a_{2n}B_n e^{-ikx}]e^{-i(\varepsilon+2neV)t/\hbar} .
    \label{eq:wfelectronhole} 
\end{align}

The wavefunctions in region $N_{R}$ produced by an electron-like quasiparticle incident from the left can be written as

\begin{align}
   \psi_{e}=\displaystyle\sum_{n} [C_ne^{ikx}+a_{2n+1} D_n e^{-ikx}]e^{-i(\varepsilon+(2n+1)eV)t/\hbar}\nonumber\\
    \psi_{h}=\displaystyle\sum_{n} [a_{2n+1}C_{n}e^{ikx}+D_{n}e^{-ikx}]e^{-i(\varepsilon+(2n+1)eV)t/\hbar},
\end{align}
where the sum over $n$ represents contributions from multiple reflections and is taken over all integers from -$\infty$ to $\infty$.

We can relate the incoming electrons to the outgoing electrons in regions $N_{L}$ and $N_{R}$ using a scattering matrix $S_e$. Similarly there will be a scattering matrix $S_h$ connecting the holes in the two regions:
\begin{align}
 \begin{pmatrix}
 B_n  \\
 C_n
 \end{pmatrix}=S_{e} \begin{pmatrix}
 a_{2n}A_n + J\delta_{n0}\\
 a_{2n+1}D_n
 \end{pmatrix},\hspace{1em}
  \begin{pmatrix}
 A_n  \\
 D_{n-1}
 \end{pmatrix}=S_{h} \begin{pmatrix}
 a_{2n}B_n \\
 a_{2n-1}C_{n-1}
 \end{pmatrix}  
\label{eq:sis:scat}
\end{align}

The scattering matrices that represent the barrier for the electrons $S_{e}$ and holes $S_{h}$ are \cite{averin1995ac} 
\begin{align}
S_{h}=S_{e}^*,\hspace{1em} S_{e}=
 \begin{pmatrix}
  r & t  \\
  t & -r^*t/t^* 
 \end{pmatrix},
 \label{eq:sis:scatmatrix_s}
\end{align}
where $r$ is the amplitude of the quasiparticle being reflected in the normal region, and $t$ is the amplitude of the quasiparticle being transmitted in the normal region. Here the transparency is $D=|t|^2$.

Eliminating the wavefunction amplitudes $C_n$ and $D_n$ in $N_R$ using \cref{eq:sis:scat} and \cref{eq:sis:scatmatrix_s}, the recursion relation for the amplitudes $A_n$ and $B_n$ will be 
\begin{align}
 A_{n+1}^{}-a_{2n+1}^{}a_{2n}^{}A_n^{}=\sqrt{R}(a_{2n+2} B_{n+1}^{}-a_{2n+1}^{} B_n^{})+Ja_1^{}\mathit{\delta}_{n0} 
 \label{sup:rec1}
\end{align}
\begin{align}
D\frac{a_{2n+1}^{} a_{2n+2}^{}}{1-(a_{2n+1}^{})^2}B_{n+1}^{}&-\bigg[D\bigg(\frac{(a_{2n+1}^{})^2}{1-(a_{2n+1}^{})^2}+\frac{(a_{2n}^{})^2}{1-(a_{2n-1}^{})^2}\bigg)+1-(a_{2n}^{})^2\bigg]B_n^{}\nonumber\\
&+D\frac{a_{2n}^{}a_{2n-1}^{}}{1-a_{2n-1}^{}}B_{n-1}^{}=-\sqrt{R}J\mathit{\delta}_{n,0},
\label{sup:rec2}
\end{align}
with a detailed derivation given in Ref~\cite{wtthesis}.

In the wavefunctions coefficients $A_n^{\rightarrow{}}$, $B_n^{\rightarrow{}}$, $C_n^{\rightarrow{}}$, $D_n^{\rightarrow{}}$, the symbol $\rightarrow{}$ is used to discern an electron-like qusiparticle incident from the left as opposed to an electron-like quasiparticle incident from the right.
To properly keep track of the left or the right interface where it is Andreev reflected from a quasiparticle incident from the right or left, we write superscript $A^{\rightarrow}$, $B^{\rightarrow}$, $A^{\leftarrow}$, $B^{\leftarrow}$, and label the Andreev reflection according to the interface. The recursion relations for $A^{\rightarrow}$, $B^{\rightarrow}$ are Eqs.~10 and 11 in the main text while the resulting recursion relations for $A^{\leftarrow}$, $B^{\leftarrow}$ are 
\begin{widetext}
\begin{align}
 A_{n+1}^{\leftarrow}-a_{2n+1}^{L}a_{2n}^{R}A_n^{\leftarrow}=\sqrt{R}(a_{2n+2}^{R} B_{n+1}^{\leftarrow}-a_{2n+1}^{L} B_n^{\leftarrow})+J^Ra_1^{L}\mathit{\delta}_{n0} 
 \label{sup:rec1a}
\end{align}
\begin{align}
D\frac{a_{2n+1}^{L} a_{2n+2}^{R}}{1-(a_{2n+1}^{R})^2}B_{n+1}^{\leftarrow}&-\bigg[D\bigg(\frac{(a_{2n+1}^{L})^2}{1-(a_{2n+1}^{L})^2}+\frac{(a_{2n}^{R})^2}{1-(a_{2n-1}^{L})^2}\bigg)+1-(a_{2n}^{R})^2\bigg]B_n^{\leftarrow}\nonumber\\
&+D\frac{a_{2n}^{R}a_{2n-1}^{L}}{1-a_{2n-1}^{L}}B_{n-1}^{\leftarrow}=-\sqrt{R}J^R\mathit{\delta}_{n0}.
\label{sup:rec2a}
\end{align}
\end{widetext} We see that the recursion relations for $A^{\leftarrow}$, $B^{\leftarrow}$ are identical to $A^{\rightarrow}$, $B^{\rightarrow}$ simply by substituting superscripts $\rightarrow$ for $\leftarrow$, $L$ for $R$, and $R$ for $L$.

\section{Subgap current for different electrode temperatures}

In Fig.~4 of the main paper, theoretical $I-V$ characteristics are plotted for different temperatures. While both electrodes of a junction are often at the same temperature, situations where there is a temperature difference are also of interest. Asymmetric junctions in temperature have been used as electronic refrigerators \cite{pekola2014refrigerator,muhonen2012micrometre,RevModPhys.78.217}. Thermal transport and heating effects can be important in small junctions. For the current work, the sharpened STM tip may have more difficulty dissipating heat and thus reach a higher temperature than the superconducting sample. We can simulate the effect of this by independently specifying the temperatures $T_{L}$ and $T_{R}$ on the left and right sides of the junction.

\begin {figure}[t]
\centering
\includegraphics[width=1\linewidth]{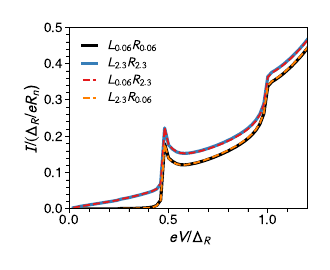}
\caption{Calculated $I-V$ characteristics for different electrode temperatures with transparency $D=0.25$, ${\Delta}_{R}=1$ meV, and ${\Delta}_{L}=0.5$ meV. The black line has $T_{L}=T_{R}=0.06$ K. The blue line has $T_{L}=T_{R}=2.3$ K. The dashed orange line lies on the black line but have $T_{L}=2.3$ K and $T_{R}=0.06$ K. The dashed red line has $T_{L}=0.06$ K and $T_{R}=2.3 $ K, and lies on top of the blue line.}
\label{temp2}
\end{figure}

\Cref{temp2} shows dc $I-V$ characteristics for four different scenarios. For all the curves, we set the superconducting gap ratio to ${\Delta}_{L}/{\Delta}_{R}=0.5$, the transparency to $D=0.25$, and ${\Delta}_{R}=1$ meV. The black line corresponds to both sides of the junction at 0.06 K, while the blue line shows $T_{L}=T_{R}=2.3$ K.  For the higher temperature, the MAR steps occur at the same voltages, but the subgap currents are larger due to a larger population of thermally excited quasiparticles. The other two cases correspond to a temperature difference: the red dashed line shows $T_{L}=0.06$ K and $T_{R}=2.3$ K, while the orange dashed line is for $T_{L}=2.3$ K  and $T_{R}=0.06$ K. Comparing the four curves shows that the behavior of the $I-V$ curve is largely governed by the temperature of the electrode with the larger superconducting gap ($T_{R}$, in this example). Thus, for the data presented in the main text, we suspect the $I-V$ curves are more strongly influenced by the temperature of the sample, which had a larger gap than the tip.

\clearpage

\newpage
\section{voltage-biased and current-biased characteristic}
\begin {figure}[t]
\centering
\includegraphics[width=1\linewidth]{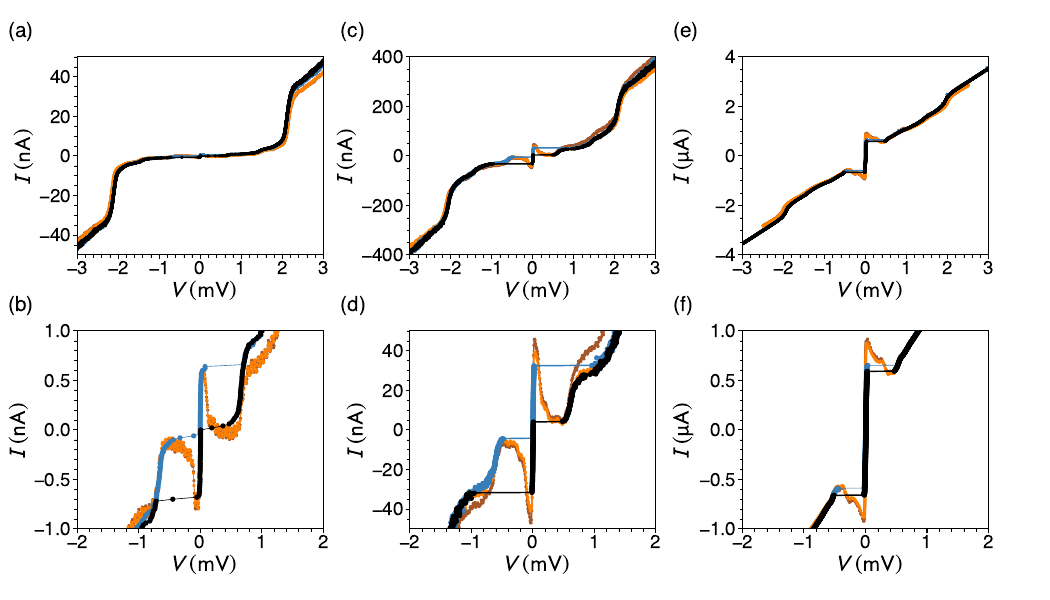}
\caption{Current-voltage characteristics for junction tunneling resistance of (a)(b) $R_n=57$ k$\Omega$, (c)(d) 7.9 k$\Omega$, and (e)(f) 1.08 k$\Omega$ measured with the voltage-biased mode and the current-biased mode. In each, the brown and orange lines show forward and reverse sweeps taken from the voltage-biased mode, and the blue and black points show forward and reverse sweeps taken from the current-biased mode.}
\label{main_supple}
\end{figure}

In the main paper, Fig.~7 shows current-biased $V(I)$ characteristics to show the detailed switching and retrapping currents. \Cref{main_supple} shows $I(V)$ and $V(I)$ characteristics taken with a voltage and current bias for junction resistance (a)(b) $R_n=57$ k$\Omega$, (c)(d) 7.9 k$\Omega$ and (e)(f) 1.08 k$\Omega$. The brown curves and the orange curves represents the forward (from negative to positive voltage) and reverse sweep (from positive to negative voltage) from the voltage-biased mode, respectively. The blue and the black data represent the forward (from negative to positive current) and reverse sweep (from positive to negative current) from the current-biased mode. We see both the voltage-biased and the current-biased curves agree with each other well except the slight current drift due to noise when the feedback is turned off to acquire the curves. Additionally, there might also be a slow vertical drift when the feedback is off so the start point of the forward sweep and the end point of the reverse sweep are offset.

\clearpage

\clearpage
\section{\didv \hspace{1pt} characteristic from the voltage-biased mode}
\begin {figure}[t]
\centering
\includegraphics[width=0.6\linewidth]{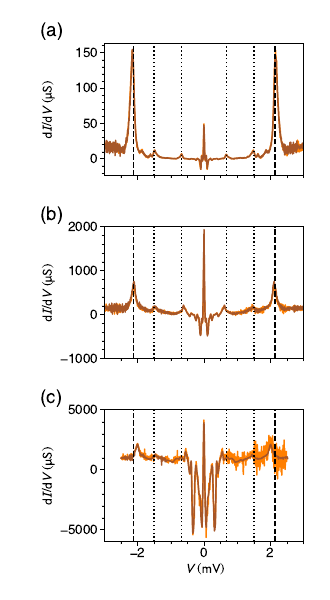}
\caption{$\didv$ characteristics for junction tunneling resistance of (a) $R_n=57$ k$\Omega$, (b) 7.9 k$\Omega$, and (c) 1.08 k$\Omega$. The brown and orange curves show forward and reverse sweeps. The black dashed lines indicate $V=\pm 2.13,\pm1.5$, and $\pm0.68$ mV. }
\label{dIdV}
\end{figure}

The location of the peaks in the $\didv$ curves from voltage-biased measurements $I(V)$ are sensitive to the superconducting gaps and transparencies. Here we show the $\didv$ curves taken in the voltage biased mode. The lock-in frequency is 0.92 kHz with a drive amplitude 10 $\mu$V. \Cref{dIdV} shows the $\didv$ characteristic for junction resistance (a) $R_n=57$ k$\Omega$, (b) 7.9 k$\Omega$ and (c) 1.08 k$\Omega$. The brown curves and the orange curves represent the forward sweep and the reverse sweep. The black dashed lines indicated indicate $V=\pm 2.13,\pm1.5$, and $\pm0.68$ mV and are at the same positions in the three plots as a guide to the eye. We can see that the peaks around the sample gaps $\mathit{\Delta}_{s}$ remain roughly the same even when the junction resistance is smaller. However the peaks around the tip gap $\mathit{\Delta}_{t}$, and sum of the gaps $\mathit{\Delta}_{t}+\mathit{\Delta}_{s}$ decrease when the junction resistance becomes smaller.

\clearpage
\section{Multi-channel fits}

\begin {figure}[t]
\centering
\includegraphics[width=1.0\linewidth]{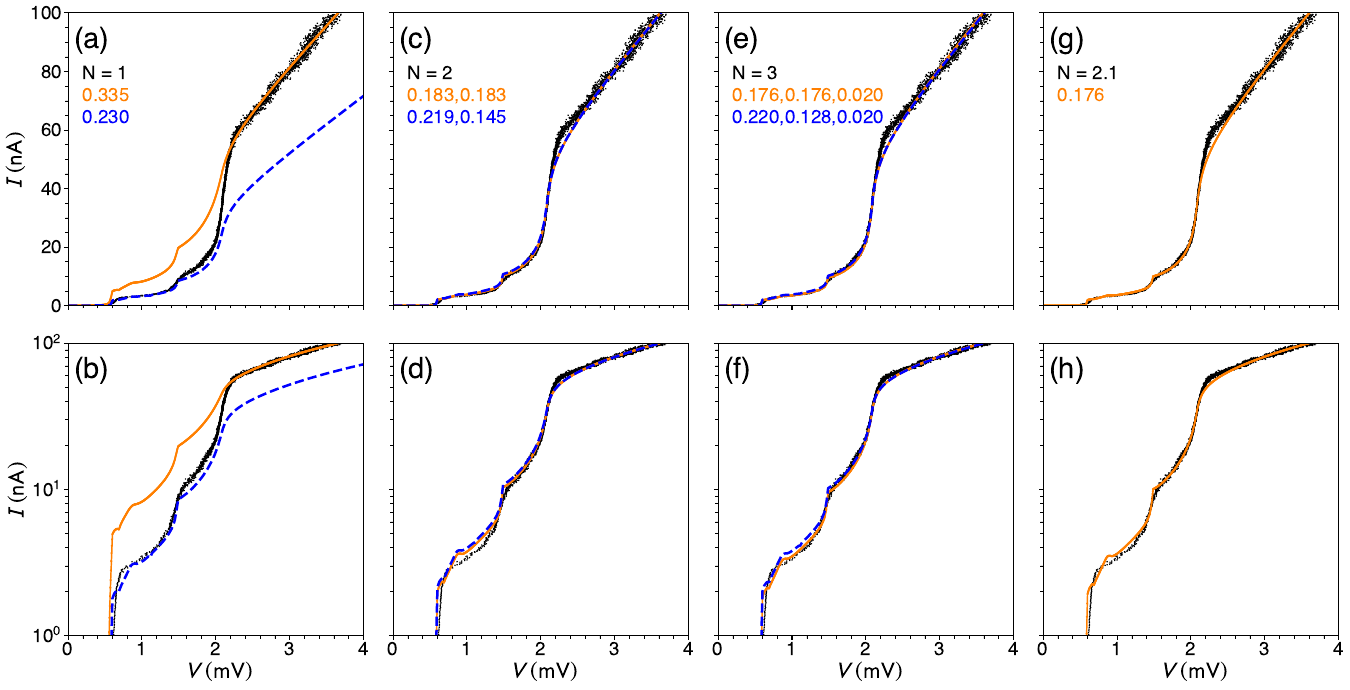}
\caption{Comparison of multi-channel MAR $I-V$ characteristics (lines) with measured values of $V(I)$ (symbols) at  $R_n = 35\ \mathrm{k}\Omega$. The calculations correspond to (a) and (b) one, (c) and (d) two, (e) and (f) three channels, with the transparencies of the individual channels listed at the top of each column. (g) and (h) Good agreement to the data is found for a non-integral number of identical channels.}
\label{multii}
\end{figure}

In this section, we motivate our decision to model the data with a non-integral number of effective conduction channels of a single transparency. In each panel of Fig. \ref{multii}, the $V(I)$ curve taken with a reverse sweep for $R_n = 35\ \mathrm{k}\Omega$ is shown. Also plotted is the multi-channel MAR current $I(V) = \displaystyle\sum_{i=1}^N I_0(D_i, \Delta_{s}, \Delta_{t}, T, V)$ calculated for different values of $N$ and $D_i$.

Figure \ref{multii} (a) and (b) show results for $N = 1$. The measured values at large $V$ are reproduced with $D = 0.335$, while much of the subgap structure is captured with $D = 0.230$. From this, it is clear that there must be multiple conduction channels, with $\sum D_i \approx 0.35$ and each $D_i < 0.230$.

In Fig. \ref{multii} (c) and (d), curves for $N = 2$ show good agreement over the full voltage range, with a slight overestimate near the voltage corresponding to the tip gap.  There is only a small difference at low voltage between identical channels and two channels with a spread in $D$.  The low-voltage features can be better reproduced by adding a third channel with significantly lower transparency, as shown in Fig. \ref{multii} (e) and (f).

Thus, we conclude that the current for this junction is predominantly carried by two channels, with identical channels producing good agreement.  To characterize the transport at this $R_n$, we assume an effective channel number  $N = 2.1$ with $D = 0.176$, as shown in Fig. \ref{multii} (g) and (h).

\clearpage

\clearpage
\section{Conductance vs retraction-distance curves for various junction resistances}
\begin {figure}[t]
\centering
\includegraphics[width=1\linewidth]{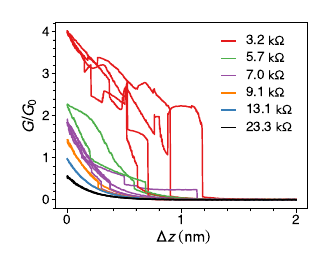}
\caption{Normalized conductance $G/G_0$ $vs$ $\Delta z$ where $G_0 = e^2/(\pi \hbar)$ is the conductance quantum and $\Delta z$ is the distance from the original tip position for different initial values of the tunnel resistance $R_n$. All measurements were made at $V_{b}=3.5$ mV and $T=1.5$ K. For $G \geq G_0$, the curves have occasional sudden jumps, suggesting the tip is forming a point contact with the surface. For $G>3$ $G_0$, the tip probably makes solid contact with the sample and the large variation in conductance may be due to different parts of the tip leaving the surface at slightly different separations.  }
\label{conductance}
\end{figure}

As we discussed in the main text, for $R_n <5$ k$\Omega$ the tip and the sample were probably in contact. To better understand this contact regime, we measured how the tunneling current at a fixed voltage bias varied with the distance of the tip from the surface. For these measurements, we set the junction resistance $R_n$ (i.e. setting the value for $I$ at a bias voltage of $V_{b}=3.5$ mV), measured the current setpoint $I$, and then found $R_n=V_{b}/I$. In the conventional tunneling limit, $I$ $\propto$ $e^{-\alpha z}$, where $z$ is the distance between the tip and the sample. Thus, setting a smaller junction resistance $R_n$ requires higher tunneling current, which results in smaller tip-sample separation. We then turned off the feedback and recorded the current $I$ while retracting the tip. \Cref{conductance} shows the conductance $G=I/V_{b}$, normalized by the conductance quantum $vs$ the retraction distance $\mathit{\Delta}z$. The measurement was repeated for six initial values of $R_n$, as labeled, although the curves were not acquired in the same order as the $R_n$ values. 

Examining \cref{conductance}, for 5 k$\Omega < R_n<  12.9$ k$\Omega$, the smallest steps are about 0.2 $G_0$, signaling the tip is very close to the sample. Two orange curves starting at $R_n$ = 9.1 k$\Omega$, one curve shows a fairly smooth exponential-like dependence while the other shows similar behavior but with some small sudden jumps in conductance. It is possible that material was picked up or a slight deformation of the tip or sample occurred during the data acquisition of the first curve. Similar behaviors are also shown in $R_n$ = 7.0 k$\Omega$, and 5.7 k$\Omega$. In contrast, the curves with $R_n > 12.9$ k$\Omega$ do not show conductance jumps. The red curves, which have an initial resistance of 3.2 k$\Omega$, show many large random jumps in conductance as the tip is withdrawn. Their behavior is probably due to different parts of the tip breaking contact with the sample at random times as the distance between the tip and sample increases. Therefore we can safely conclude that for $R_n<$ 5 k$\Omega$, the tip is making solid physical contact with the sample, causing deformation of the tip or sample, or the transfer of material between the tip and sample while retracting the tip from the sample. This result is comparable to behavior seen previously in break junctions, where the typical variations in conductance are of the order of $G_0$ \cite{scheer2001proximity,ludoph2000multiple,scheer1998signature,scheer1997conduction,agrait2003quantum}.

\newpage

\begin{table}
\caption{symbol for different parameters}
 \centering
\begin{ruledtabular}
\begin{tabular}{cc}
Switching current & $I_s$ \\
Retrapping current & $I_r$ \\
Critical current & $I_c$ \\
Quasiparticle current & $I_{qp}$ \\
Josephson energy & $E_J$ \\
Coulomb charging energy & $E_c$ \\
Plasma energy  & $\hbar w_p$ \\
Stewart-McCumber parameter & $\beta_c$
\end{tabular}
\end{ruledtabular}
\end{table}



%

\end{document}